\documentclass[12pt,preprint]{aastex}

\def\etal{{et\thinspace al.}\ }               

\newcommand{\E}[1]{\,10^{#1}}                              
\newcommand{\ET}[1]{\times 10^{#1}}                  

\begin{document}

\title{Empirical diagnostics of the starburst-AGN connection}

\author{R. Cid Fernandes\altaffilmark{1}}
\affil{Department of Physics \& Astronomy, Johns Hopkins University,
3400 N. Charles St., Baltimore, MD, 21218}
\email{cid@pha.jhu.edu}

\author{T. Heckman}
\affil{Department of Physics \& Astronomy, Johns Hopkins University,
3400 N. Charles St., Baltimore, MD, 21218}
\email{heckman@pha.jhu.edu}

\author{H. Schmitt\altaffilmark{2}}
\affil{Space Telescope Science Institute, 3700 San
Martin Drive, Baltimore, MD, 21218 \and National Radio Astronomy Observatory, PO Box 0, Socorro, NM 87801}
\email{hschmitt@nrao.edu}

\author{R. M. Gonz\'alez Delgado}
\affil{Instituto de Astrof\'{\i}sica de Andaluc\'{\i}a (CSIC) , Apto.\
3004, 18080 Granada, Spain}
\email{rosa@iaa.es}

\and

\author{T. Storchi-Bergmann}
\affil{Instituto de F\'{\i}sica, Universidade Federal do Rio Grande do
Sul, C.P 15001, 91501-970, Porto Alegre, RS, Brazil}
\email{thaisa@if.ufrgs.br}

\altaffiltext{1}{Gemini Fellow. On leave of absence from Depto.\ de
F\'{\i}sica - CFM, UFSC, Florian\'opolis, SC, Brazil}

\altaffiltext{2}{Jansky Fellow}

\date{\today}

\begin{abstract}

We examine a representative sample of 35 Seyfert 2 nuclei.  Previous work
has shown that nearly half (15) of these nuclei show the direct (but
difficult-to-detect) spectroscopic signature at optical/near-UV wavelengths
of the hot massive stars that power circum-nuclear starbursts. In the
present paper we examine a variety of more-easily-measured quantities for
this sample, such as the equivalent widths of strong absorption features,
continuum colors, emission-line equivalent widths, emission line ratios and
profiles, far-IR luminosities, and near-UV surface brightness. We compare
the composite starburst $+$ Seyfert 2 nuclei to ``pure'' Seyfert 2 nuclei,
Starburst galaxies and normal galactic nuclei.  Our goals are to verify
whether the easily-measured properties of the composite nuclei are
consistent with the expected impact of a starburst, and to investigate
alternative less-demanding methods to infer the presence of starbursts in
Seyfert 2 nuclei, applicable to larger or more distant samples.  We show
that starbursts do indeed leave clear and easily quantifiable imprints on
the near-UV to optical continuum and emission line properties of Seyfert
2's. Composite starburst $+$ Seyfert 2 systems can be recognized by: (1) a
strong ``Featureless Continuum'' (FC), which dilutes the CaII K line from
old stars in the host's bulge to an equivalent width $W_K < 10$ \AA; (2)
emission lines whose equivalent widths are intermediate between Starburst
galaxies and ``pure'' Seyfert 2's; (3) relatively low excitation
line-ratios, which indicate that part of the gas ionization in these
Seyfert 2's (typically $\sim 50\%$ of H$\beta$) is due to photoionization
by OB stars; (4) large far IR luminosities ($\ga 10^{10}$ L$_\odot$); (5)
High near-UV surface brightness ($\sim 10^3$ L$_{\odot}$ pc$^{-2}$).  These
characteristics are all consistent with the expected impact of
circum-nuclear starbursts on the observed properties of Seyfert
2's. Furtheremore, they offer alternative empirical diagnostics of the
presence of circum-nuclear starbursts from a few easily measured
quantities.

\end{abstract}


\section{Introduction}

\label{sec:Introduction}

The role of starbursts in active galactic nuclei (AGN) is an issue which
permeates the history of AGN literature.  Theoretical scenarios for a
``starburst-AGN connection'' vary from those in which starbursts are a key
piece of the AGN machinery, either through symbiotic or evolutionary
processes (e.g., Perry \& Dyson 1985; Sanders \etal 1988; Norman \&
Scoville 1988; Rees 1989; Daly 1990; Terlevich \etal 1992; Williams, Baker
\& Perry 1999), to those in which these are fundamentally decoupled
phenomena, but which are likely to coexist merely because both are
triggered by and live on gas fueling (e.g., Byrd \etal 1986; Byrd,
Sundeluis \& Valtonen 1987; Lin, Pringle \& Rees 1988; Heller \& Shlosman
1994; Hernquist \& Mihos 1995; Mihos \& Hernquist 1996). Given the enormous
breadth of possibilities in between these extremes, it is likely that
progress in understanding this connection will be driven more by input from
observations than from theoretical considerations.

Observational work during the past decade has indeed substantially
advanced our ability to recognize starbursts in the complex inner
environment of AGN. Optical and UV studies of Seyfert 2 galaxies have
been particularly enlightening in this respect.  In a series of papers
(Heckman \etal 1997; Cid Fernandes, Storchi-Bergmann \& Schmitt 1998;
Gonz\'alez Delgado \etal 1998; Storchi-Bergmann, Cid Fernandes \&
Schmitt 1998; Schmitt, Storchi-Bergmann \& Cid Fernandes 1999;
Storchi-Bergmann \etal 2000; Gonz\'alez Delgado, Heckman \& Leitherer
2001) we have detected unambiguous signatures of young massive stars
within $\sim 300$ pc of the nucleus in 30 to 50\% of Seyfert 2's by
means of high quality optical and, whenever possible, UV
spectroscopy. This result fits well with the near IR studies of Oliva
\etal (1995, 1999), which find a comparable incidence of starbursts in
Seyfert 2's through measurements of the stellar mass-to-light ratio.
The starbursts in these composite
starburst$+$AGN systems can make a significant contribution to the
total luminosity output.  In fact, studies specifically aimed at
candidate composite galaxies, as selected by emission line ratios
intermediate between HII regions and AGN on classical diagnostic
diagrams, demonstrate that their far IR and radio properties are
dominated by star-forming activity at the 90\% level (Hill \etal 2001,
1999).  Even the presence of compact radio cores, a more classical
indicator of AGN activity (Condon \etal 1991; Sramek \& Weedman 1986;
Norris \etal 1990), can be accounted for by
star-formation (Hill \etal 2001; Kewley \etal 2000; Smith \etal 1998a,b;
Lonsdale, Smith \& Lonsdale 1993). Other recent results are covered in the
review by Veilleux (2001) and the volume edited by Aretxaga, Kunth \&
Mugica (2001).

Evidence is also steadily accumulating that associates AGN with
star-formation on galactic scales.  The ubiquity of super-massive
black holes in the nuclei of normal galaxies in the local universe (Ho
1999), the proportionality between bulge and black-hole masses
(Magorrian \etal 1998; Gebhart \etal 2000; Ferrarese \& Merrit 2000),
the link between $M_{black-hole} / M_{bulge}$ and the age of the last
major star-formation episode in the spheroid (Merrifield, Forbes \&
Terlevich 2000), all imply that the creation of these black holes and
the ensuing QSO activity were an integral part of the formation of
ellipticals and galactic bulges. This fossil evidence indirectly
traces a much more active past, in which both copious star-formation
and nuclear activity coexisted.  Since the technology to directly
study this high redshift era in detail is not available, we must guide
our study of the starburst-AGN connection by data gathered on nearby
objects in which traces of this connection are caught {\it in
fraganti}. This means refining our knowledge of circum-nuclear
starbursts in Seyfert galaxies. In the optical-UV range, as hinted
above, this is best accomplished studying type 2 Seyferts. Their
favorable geometry, in which the blinding glare of the nucleus is
blocked away by a dusty torus, facilitates the detection of features
from circum-nuclear starbursts, thus making them the best-suited local
laboratories to study the starburst-AGN connection.

Advancing our empirical understanding of circum-nuclear starbursts in
Seyfert galaxies can be described as a 3-stage process: (I) Identifying
starbursts in these systems; (II) characterizing the properties of the
starburst (age and mass, or star-formation rate) and the AGN (black-hole
mass, accretion rate); and (III) investigating possible connections between
these properties.  Once we get to this third stage fundamental questions
can start to be addressed. For instance: Is star-formation inextricably
associated with AGN activity?  Are the starburst and AGN powers related,
say, by a $M_{starburst} \propto M_{black-hole}$ proportionality between
their masses?  Or, analogously, is the star-formation rate connected to the
rate of accretion onto the nuclear super-massive black-hole? Does the
central engine evolve in parallel to the starburst around it? Much work
remains to be done before tackling these key questions.

The work on Seyfert 2's reviewed above pertains mostly to stage I of
this process, as it essentially demonstrates how certain spectral
features of massive stars can be used as sign-posts of starburst
activity. Estimates of the recent history of star-formation in the
circum-nuclear regions were also presented, thus touching the domain
of stage II.  These estimates come associated with known difficulties
in accurately retrieving the detailed star-formation history of
stellar systems (e.g., Leitherer 1999; Cid Fernandes \etal 2001;
Kennicutt 1998), but they represent the current state-of-the-art
regarding the characterization of the basic starburst properties.

This paper deals mainly with stage I, that is, the identification of
starbursts in Seyfert 2's, using the data sets of Cid Fernandes \etal
(1998) and Gonz\'alez Delgado \etal (2001, hereafter GD01) as a
training set.  This is probably the largest sample whose optical
spectra have been systematically and carefully screened for starburst
features, so it represents a good starting point. The analysis is
geared towards verifying whether composite Seyfert 2/starburst
systems, those in which stellar features produced by young populations
have been directly detected, exhibit other properties which
differentiate them from systems not showing signs of starburst
activity (``pure'' Seyfert 2's).  Unlike the detailed analysis
reported by the previous work on these data, we here limit ourselves
to easily measurable quantities.  Specifically, we will use equivalent
widths of strong absorption bands and continuum colors in the near UV,
emission line equivalent widths, fluxes and line ratios, far IR
luminosities and near UV surface brightness. The purpose of this
analysis is two-fold:

\begin{itemize}

\item[(1)] To verify whether the optical continuum and emission line
properties of composite systems are consistent with the expected
impact of starbursts upon these observables.

\item[(2)] To investigate alternative ways to infer the presence of
circum-nuclear starbursts in Seyfert 2's that may provide more
straightforward diagnostics applicable to larger and/or more distant
samples.

\end{itemize}

The first of these points serves both as a consistency check and to give a
sense of the role played by starbursts in the phenomenology and energetics
of AGN. The relevance of the second point is that increasing the statistics
is crucial to improve upon the still sketchy overall census of
circum-nuclear starbursts in Seyfert 2's, to extend these studies to higher
redshifts and to provide the raw material for further stage II and III
studies.

In \S\ref{sec:Sample} we describe the data for the Seyfert 2's and
comparison samples used in this study.  In \S\ref{sec:Synthesis} we deal
with the problem of estimating the ``Featureless Continuum'' (FC)
strength. We apply a method based on population synthesis, but which can be
parameterized in terms of a few easily measurable quantities. Essentially,
we are able to estimate the FC contribution to the optical light with
little more than the equivalent width of the CaII K absorption line, a
method which may be of wide applicability in Seyfert 2 studies. In
\S\ref{sec:Emission_Lines} we present an analysis of the emission line
equivalent widths, line ratios and profiles, and discuss how they are
affected by the presence of circum-nuclear starbursts. Emission line-FC
correlations are also presented and discussed. In \S\ref{sec:IRAS} we
investigate how the composite systems behave in terms of far IR luminosity
and how they fare in comparison to normal and interacting galaxies.  In
\S\ref{sec:SurfBrightness} we compare the near UV surface brightness
of Seyfert 2's in our sample to that of normal and Starburst galaxies.
In \S\ref{sec:Discussion} we collect the results obtained in a set of
empirical criteria which may be used to diagnose the presence of
circum-nuclear starbursts in Seyfert 2's. We also discuss the meaning
of the composite/``pure'' Seyfert 2 classification and present
tentative evidence of evolutionary effects in our sample.  Section
\ref{sec:Conclusions} summarizes our conclusions.

\section{The Database}

\label{sec:Sample}

In this study we merge the Seyfert 2's in the southern sample studied by
Cid Fernandes \etal (1998) with those in the northern sample of GD01,
totaling 35 galaxies. Details of the observations, sample selection and
previous analysis of these data can be found in the original papers and
Heckman \etal (1997), Gonz\'alez Delgado \etal (1998), Storchi-Bergmann
\etal (1998), Schmitt \etal (1999) and
Storchi-Bergmann \etal (2000, hereafter SB00). We concentrate our analysis
on the nuclear spectra, extracted through $1.2^{\prime\prime}
\times 2.1^{\prime\prime}$ apertures for the northern sample and
$2^{\prime\prime} \times 2^{\prime\prime}$ for the southern one. At the
distances of our galaxies this covers a region 60--860 pc in radius, with a
median of 300 pc ($H_0 = 75$ km$\,$s$^{-1}\,$Mpc$^{-1}$ is adopted
throughout this paper).  We will also focus on the near UV to optical
region of the spectrum, between 3500 and 5100 \AA.

These 35 galaxies will hereafter be referred to as the ``Seyfert 2
sample''. They constitute a good data set to investigate the
starburst-AGN connection because the above papers have scrutinized
these sources in search for signs of a starburst
component. Specifically, the following signatures of a starburst were
examined: (1) the presence of far-UV stellar wind lines (NV
$\lambda$1240, SiIV$\lambda$1397 and CIV$\lambda$1549) due to O stars;
(2) high order Balmer absorption lines of HI and HeI in the near UV;
(3) the WR bump underneath HeII$\lambda$4686. The far-UV features are
the cleanest signatures of recent star formation, but, unfortunately,
there are only a handful of Seyfert 2's for which HST far-UV
spectroscopy is feasible. All four galaxies in the sample for which we
acquired such data (Mrk 477, NGC 5135, NGC 7130 and IC 3639) have
their far-UV spectrum dominated by young stars. For the remaining
galaxies, detection of a starburst component relies on either high
order Balmer lines, which originate in the photosphere of O, B and A
stars, or the WR bump, a tracer of very recent (3--6 Myr) star
formation. Although more subtle than the far-UV features, these are
reliable and more easily accessible indicators of the presence of a
starburst. This is confirmed by the fact that either of these two
features is also detected in the galaxies for which we have identified
stellar wind lines in the far-UV.  The least certain of these
diagnostics is the WR bump, because independent evidence of the
presence of WR stars is hard to obtain (GD01; Shaerer
2001). Nevertheless, this is the simplest interpretation of this
feature, given the similarity to the feature sometimes seen in
star-forming galaxies (``WR galaxies''), the lack of convincing
alternatives and the simultaneous detection of other starburst
features in some of the sources.

The information to be used in this paper comprises fluxes and
equivalent widths ($W$) of the [OII]$\lambda\lambda$3726,3729,
HeII$\lambda$4686, H$\beta$ and [OIII]$\lambda$5007 emission lines;
the $W$'s of CaIIK$\lambda$3933 ($W_K$), CN$\lambda$4200 ($W_{CN}$)
and G-band$\lambda$4301 ($W_G$) absorption bands, the continuum fluxes
at 3660, 4020 and 4510 \AA, as well as the 12 $\mu$m and far IR
luminosities as measured by IRAS. These are listed in
Tables~\ref{tab:TABELAO1} and \ref{tab:TABELAO2}, along with
heliocentric velocities extracted from NED\footnote{The NASA/IPAC
Extragalactic Database (NED) is operated by the Jet Propulsion
Laboratory, California Institute of Technology, under contract with
the National Aeronautics and Space Administration.}.
Table~\ref{tab:TABELAO2} also lists the H$\alpha$/H$\beta$ ratio for
the Seyfert 2 sample, mostly compiled from the literature, since the
Kitt Peak spectra do not cover the H$\alpha$ region. These
measurements come from spectra obtained with apertures larger than
those used in our observations. For this reason, and also because of
the intrinsic uncertainties associated with reddening corrections, we
will concentrate on reddening insensitive diagnostics inasmuch as
possible.  Galaxies which had hidden Seyfert 1 nuclei revealed through
spectropolarimetry are indicated by the corresponding reference in the
last column of Table~\ref{tab:TABELAO2}.

All absorption line $W$'s and continuum fluxes used in this paper were
measured according to the system explained in Cid Fernandes \etal
(1998). Emission line and continuum fluxes were corrected for Galactic
extinction following the extinction law of Cardelli, Clayton \& Mathis
(1989, with $R_V = 3.1$), and $A_B$ values from Schlegel, Finkbeiner
\& Davis (1998) as listed in NED.

\begin{deluxetable}{lrrrrrrrrrrrrrr}
\rotate
\tabletypesize{\tiny}
\tablecaption{Seyfert 2 sample: Continuum, absorption and emission line properties}
\tablewidth{0pc}
\tablehead{
\colhead{Galaxy}&
\colhead{$W_K$}&
\colhead{$W_{CN}$}&
\colhead{$W_G$}&
\colhead{$\frac{F_{3660}}{F_{4020}}$}&
\colhead{$\frac{F_{4510}}{F_{4020}}$}&
\colhead{$W_{[OII]}$}&
\colhead{$W_{HeII}$}&
\colhead{$W_{H\beta}$}&
\colhead{$W_{[OIII]}$}&
\colhead{$F_{[OII]}$}&
\colhead{$F_{HeII}$}&
\colhead{$F_{[OIII]}$}&
\colhead{$F_{H\beta}$}\cr
\colhead{(1)}&
\colhead{(2)}&
\colhead{(3)}&
\colhead{(4)}&
\colhead{(5)}&
\colhead{(6)}&
\colhead{(7)}&
\colhead{(8)}&
\colhead{(9)}&
\colhead{(10)}&
\colhead{(11)}&
\colhead{(12)}&
\colhead{(13)}&
\colhead{(14)}&
}
\startdata
CGCG420-015           		 & 14.0 &  9.7 &  8.4 & 0.53 & 1.67 &  32.5 &   4.7 &  16.9 & 208.5 &   7.9$\E{-15}$ &   3.6$\E{-15}$ &   1.7$\E{-13}$ &   1.4$\E{-14}$ \cr
ESO 417-G6            		 & 14.6 & 11.3 &  9.4 & 0.57 & 1.29 &  42.1 &       &   2.8 &  18.1 &   1.8$\E{-14}$ &               &   2.3$\E{-14}$ &   2.8$\E{-15}$ \cr
ESO 362-G8\tablenotemark{\star} &  7.3 &  6.6 &  5.9 & 0.35 & 1.21 &   3.2 &       &       &  13.1 &   4.2$\E{-15}$ &               &   5.3$\E{-14}$ &           \cr
Fairall 316           		 & 17.2 & 14.2 & 11.3 & 0.61 & 1.59 &  30.1 &       &   1.4 &  22.2 &   8.8$\E{-15}$ &               &   2.0$\E{-14}$ &   1.2$\E{-15}$ \cr
IC 1816             	  	 & 11.8 & 10.0 &  8.2 & 0.65 & 1.29 &  59.8 &   4.7 &  15.7 & 244.7 &   3.0$\E{-14}$ &   4.9$\E{-15}$ &   2.6$\E{-13}$ &   1.8$\E{-14}$ \cr
IC 3639\tablenotemark{\star} 	 &  8.0 &  4.6 &  5.4 & 0.67 & 1.07 &  40.3 &   2.8 &  16.2 & 122.6 &   4.4$\E{-14}$ &   4.7$\E{-15}$ &   2.1$\E{-13}$ &   2.7$\E{-14}$ \cr
IRAS 11215-2806       		 & 13.1 &  7.5 &  7.3 & 0.54 & 1.28 &  19.9 &   1.4 &   2.9 &  52.9 &   6.1$\E{-15}$ &   1.1$\E{-15}$ &   4.3$\E{-14}$ &   2.2$\E{-15}$ \cr
MCG -05-27-013        		 & 15.4 & 10.9 &  9.7 & 0.71 & 1.54 & 114.2 &   7.2 &  22.9 & 244.8 &   1.2$\E{-14}$ &   2.0$\E{-15}$ &   8.1$\E{-14}$ &   7.0$\E{-15}$ \cr
Mrk 1\tablenotemark{\star}   	 &  7.6 &  6.0 &  4.4 & 0.83 & 1.11 & 131.4 &  12.2 &  53.5 & 765.9 &   3.4$\E{-14}$ &   4.8$\E{-15}$ &   2.7$\E{-13}$ &   2.2$\E{-14}$ \cr
Mrk 3                 		 & 11.3 & 10.2 &  6.7 & 1.04 & 1.38 & 263.4 &  12.3 &  67.8 & 752.7 &   4.9$\E{-13}$ &   3.6$\E{-14}$ &   3.0$\E{-12}$ &   2.1$\E{-13}$ \cr
Mrk 34                		 &  9.9 &  6.7 &  7.0 & 0.81 & 1.20 & 157.2 &  11.2 &  42.1 & 475.0 &   2.7$\E{-14}$ &   2.7$\E{-15}$ &   1.0$\E{-13}$ &   1.0$\E{-14}$ \cr
Mrk 78\tablenotemark{\star}  	 &  9.6 &  7.6 &  6.2 & 0.48 & 1.25 &  96.0 &   3.3 &   9.3 & 131.2 &   2.5$\E{-14}$ &   2.3$\E{-15}$ &   9.0$\E{-14}$ &   6.6$\E{-15}$ \cr
Mrk 273\tablenotemark{\star} 	 &  5.6 &  5.2 &  3.9 & 0.58 & 1.03 &  46.4 &       &  11.9 &  35.8 &   7.6$\E{-15}$ &               &   7.9$\E{-15}$ &   2.8$\E{-15}$ \cr
Mrk 348             		 & 12.5 &  8.4 &  7.7 & 0.76 & 1.30 & 118.6 &   3.7 &  22.1 & 274.7 &   4.7$\E{-14}$ &   2.8$\E{-15}$ &   2.2$\E{-13}$ &   1.7$\E{-14}$ \cr
Mrk 463\tablenotemark{\star} 	 &  3.0 &  5.6 &  0.9 & 1.05 & 1.04 & 135.0 &  11.6 &  88.1 & 978.5 &   5.5$\E{-14}$ &   4.7$\E{-15}$ &   3.3$\E{-13}$ &   3.8$\E{-14}$ \cr
Mrk 477\tablenotemark{\star} 	 &  0.3 &  2.4 &  0.2 & 1.10 & 0.98 & 168.1 &  13.2 &  92.5 & 743.2 &   7.6$\E{-14}$ &   6.4$\E{-15}$ &   4.5$\E{-13}$ &   4.5$\E{-14}$ \cr
Mrk 533\tablenotemark{\star} 	 &  4.1 &  3.1 &  3.1 & 0.68 & 1.06 &  49.8 &   9.2 &  36.3 & 429.2 &   2.2$\E{-14}$ &   6.1$\E{-15}$ &   2.2$\E{-13}$ &   2.1$\E{-14}$ \cr
Mrk 573               		 & 14.1 & 11.8 &  9.0 & 0.67 & 1.44 &  83.2 &   9.2 &  20.6 & 282.7 &   4.9$\E{-14}$ &   1.1$\E{-14}$ &   3.1$\E{-13}$ &   2.5$\E{-14}$ \cr
Mrk 607               		 & 13.8 & 12.3 &  9.9 & 0.61 & 1.43 &  19.1 &   4.0 &   7.1 &  78.1 &   8.6$\E{-15}$ &   4.5$\E{-15}$ &   9.5$\E{-14}$ &   8.0$\E{-15}$ \cr
Mrk 1066\tablenotemark{\star}	 &  5.6 &  4.3 &  3.5 & 0.75 & 1.03 &  43.9 &   2.3 &  24.3 &  93.1 &   5.9$\E{-14}$ &   4.1$\E{-15}$ &   1.6$\E{-13}$ &   4.2$\E{-14}$ \cr
Mrk 1073\tablenotemark{\star}	 &  5.7 &  5.4 &  4.7 & 0.74 & 1.06 &  40.2 &   5.1 &  19.9 & 122.1 &   3.2$\E{-14}$ &   5.7$\E{-15}$ &   1.3$\E{-13}$ &   2.2$\E{-14}$ \cr
Mrk 1210\tablenotemark{\star}	 &  7.5 &  7.6 &  5.5 & 0.83 & 1.20 &  59.3 &   8.8 &  57.0 & 614.0 &   4.1$\E{-14}$ &   9.5$\E{-15}$ &   5.6$\E{-13}$ &   5.5$\E{-14}$ \cr
NGC 1068              		 &  7.4 &  7.7 &  5.1 & 0.72 & 1.26 &  33.4 &  11.6 &  35.0 & 343.7 &   3.2$\E{-13}$ &   1.9$\E{-13}$ &   7.3$\E{-12}$ &   5.6$\E{-13}$ \cr
NGC 1358              		 & 17.8 & 16.1 & 11.5 & 0.54 & 1.50 &  17.8 &       &   0.9 &  19.7 &   9.4$\E{-15}$ &               &   3.2$\E{-14}$ &   1.4$\E{-15}$ \cr
NGC 1386              		 & 13.2 &  9.0 &  8.1 & 0.54 & 1.56 &  25.5 &   2.5 &   3.4 &  58.8 &   2.3$\E{-14}$ &   6.8$\E{-15}$ &   1.6$\E{-13}$ &   9.2$\E{-15}$ \cr
NGC 2110              		 & 13.5 & 12.8 &  8.9 & 0.78 & 1.31 & 177.6 &   2.0 &  15.8 &  78.3 &   8.6$\E{-14}$ &   1.9$\E{-15}$ &   8.1$\E{-14}$ &   1.6$\E{-14}$ \cr
NGC 3081              		 & 13.7 & 10.5 &  8.7 & 0.69 & 1.38 &  36.6 &   4.9 &   9.3 & 128.0 &   7.9$\E{-15}$ &   2.3$\E{-15}$ &   6.6$\E{-14}$ &   4.7$\E{-15}$ \cr
NGC 5135\tablenotemark{\star}	 &  2.5 &  1.7 &  2.5 & 0.77 & 0.87 &  20.6 &   1.8 &  13.2 &  51.3 &   4.5$\E{-14}$ &   4.1$\E{-15}$ &   1.1$\E{-13}$ &   2.7$\E{-14}$ \cr
NGC 5643\tablenotemark{\star}	 &  9.4 &  4.2 &  4.6 & 0.50 & 1.24 &  56.4 &   2.8 &   8.7 & 115.0 &   5.7$\E{-14}$ &   6.8$\E{-15}$ &   2.7$\E{-13}$ &   1.9$\E{-14}$ \cr
NGC 5643\tablenotemark{\star}    &  9.4 &  4.2 &  4.6 & 0.50 & 1.24 &  56.4 &   2.8 &   8.7 & 115.0 &   3.0$\E{-13}$ &   3.5$\E{-14}$ &   1.4$\E{-12}$ &   1.0$\E{-13}$ \cr
NGC 5929              		 & 14.3 & 11.0 &  9.6 & 0.71 & 1.39 & 107.0 &   1.6 &  14.5 &  58.5 &   3.9$\E{-14}$ &   1.2$\E{-15}$ &   5.0$\E{-14}$ &   1.2$\E{-14}$ \cr
NGC 6300              		 & 15.9 &  4.0 &  5.8 & 0.49 & 1.64 &  15.9 &       &       &  27.9 &   3.3$\E{-15}$ &               &   2.4$\E{-14}$ &           \cr
NGC 6890              		 & 12.6 &  1.0 &  4.8 & 0.72 & 1.42 &  16.5 &   2.4 &   3.9 &  74.2 &   4.5$\E{-15}$ &   1.3$\E{-15}$ &   4.4$\E{-14}$ &   2.2$\E{-15}$ \cr
NGC 7130\tablenotemark{\star}	 &  3.7 &  2.2 &  2.4 & 0.73 & 1.00 &  24.8 &   2.3 &  14.8 & 104.0 &   1.4$\E{-14}$ &   1.8$\E{-15}$ &   7.3$\E{-14}$ &   1.1$\E{-14}$ \cr
NGC 7212             		 & 10.4 &  8.1 &  6.9 & 0.92 & 1.33 & 211.2 &  13.8 &  55.2 & 725.5 &   7.0$\E{-14}$ &   7.0$\E{-15}$ &   3.5$\E{-13}$ &   3.0$\E{-14}$ \cr
NGC 7582\tablenotemark{\star} 	 &  3.7 &  3.1 &  3.0 & 0.55 & 1.23 &  19.2 &   1.6 &  15.2 &  33.0 &   1.3$\E{-14}$ &   2.6$\E{-15}$ &   5.6$\E{-14}$ &   2.4$\E{-14}$ \cr
\enddata
\tablenotetext{\star}{Confirmed starburst/Seyfert 2 composite}
\tablecomments{
Absorption line, continuum and emission line properties for the Seyfert 2
Sample, all extracted through small apertures ($1.2^{\prime\prime} \times
2.1^{\prime\prime}$ or $2^{\prime\prime} \times 2^{\prime\prime}$). All
equivalent widths are given in \AA\ and line fluxes in
erg$\,$s$^{-1}\,$cm$^{-2}$ (after correction for Galactic
extinction). Empty slots correspond to weak lines, which were not
measured. The flux calibration for NGC 5643, NGC 6300 and NGC 6890 are
very uncertain due to weather conditions.}
\label{tab:TABELAO1}
\end{deluxetable}

\begin{deluxetable}{lrrrrrrrr}
\tabletypesize{\tiny}
\tablecaption{Seyfert 2 sample: Additional data}
\tablewidth{0pc}
\tablehead{
\colhead{Galaxy}&
\colhead{$A_{B,Gal}$}&
\colhead{$cz$}&
\colhead{H$\alpha$/H$\beta$}&
\colhead{$\log L_{12\mu}$}&
\colhead{$\log L_{FIR}$}&
\colhead{$\log L_{H\beta}$}&
\colhead{Ref.}&
\colhead{Ref.}\cr
\colhead{(1)}&
\colhead{(2)}&
\colhead{(3)}&
\colhead{(4)}&
\colhead{(5)}&
\colhead{(6)}&
\colhead{(7)}&
\colhead{(8)}&
\colhead{(9)}\cr
}
\startdata
CGCG420-015      		& 0.37 &  8811 &  5.5 & 10.43 &   9.76 &  40.37  & dG92 &  \cr
ESO 417-G6       		& 0.09 &  4884 &  4.7 &  9.29 &   8.45 &  39.16  & S98  &  \cr
ESO 362-G8\tablenotemark{\star} & 0.14 &  4785 &      &  9.34 &   9.16 &         &      &  \cr
Fairall 316      		& 0.60 &  4950 &  6.4 &       &        &  38.80  & S98  &  \cr
IC 1816          		& 0.12 &  5080 &  6.0 &  9.75 &  10.03 &  39.98  & dG92 &  \cr
IC 3639\tablenotemark{\star}    & 0.30 &  3275 &  6.1 &  9.98 &  10.35 &  39.79  & dG92 & H97  \cr
IRAS 11215-2806  		& 0.31 &  4047 &  3.8 &  9.56 &   9.08 &  38.88  & dG92 &  \cr
MCG -05-27-013   		& 0.28 &  7162 &  5.0 &  9.75 &   9.99 &  39.88  & dG92 &  \cr
Mrk 1\tablenotemark{\star}      & 0.26 &  4780 &  5.9 & 10.47 &  10.18 &  40.03  & M94  &  \cr
Mrk 3         			& 0.81 &  4050 &  6.6 & 10.21 &  10.18 &  40.86  & M94  & MG90, T95 \cr
Mrk 34        			& 0.04 & 15140 & 10.5 & 10.33 &  10.66 &  40.69  & M94  &  \cr
Mrk 78\tablenotemark{\star}     & 0.15 & 11137 &  6.5 & 10.34 &  10.39 &  40.24  & M94  &  \cr
Mrk 273\tablenotemark{\star}    & 0.04 & 11326 &  9.3 & 10.62 &  11.84 &  39.88  & M94  &  \cr
Mrk 348       			& 0.29 &  4507 &  6.0 &  9.94 &   9.84 &  39.86  & M94  & MG90, T95 \cr
Mrk 463E\tablenotemark{\star}   & 0.13 & 14895 &  5.6 & 11.19 &  11.07 &  41.25  & M94  & MG90, T95 \cr
Mrk 477\tablenotemark{\star}    & 0.05 & 11332 &  5.4 & 10.35 &  10.67 &  41.09  & M94  & T95 \cr
Mrk 533\tablenotemark{\star}    & 0.25 &  8713 &  5.0 & 10.85 &  11.08 &  40.54  & M94  &  MG90, T95, H97\cr
Mrk 573       			& 0.10 &  5174 &  4.2 &  9.66 &   9.88 &  40.16  & M94  & K94 \cr
Mrk 607       			& 0.20 &  2716 &  4.9 &  9.53 &   9.63 &  39.10  & D88  &  \cr
Mrk 1066\tablenotemark{\star}   & 0.57 &  3605 &  8.5 &  9.90 &  10.56 &  40.06  & M94  & \cr
Mrk 1073\tablenotemark{\star}   & 0.69 &  6998 &  6.3 & 10.48 &  11.04 &  40.35  & M94  &  \cr
Mrk 1210\tablenotemark{\star}   & 0.13 &  4046 &  5.2 & 10.05 &   9.85 &  40.28  & B99  & T95 \cr
NGC 1068      			& 0.14 &  1137 &  4.5 & 10.85 &  10.79 &  40.18  & M94  & AM85 \cr
NGC 1358      			& 0.27 &  4028 &  3.4 &  8.95 &   9.34 &  38.69  & M94  &  \cr
NGC 1386      			& 0.05 &   868 &  4.9 &  8.71 &   9.09 &  38.17  & M94  &  \cr
NGC 2110      			& 1.62 &  2335 &  8.1 &  9.42 &   9.79 &  39.28  & M94  &  \cr
NGC 3081      			& 0.24 &  2385 &  4.5 &       &        &  38.75  & M94  & M00 \cr
NGC 5135\tablenotemark{\star}   & 0.26 &  4112 &  6.1 & 10.17 &  10.93 &  39.99  & M94  &  \cr
NGC 5643\tablenotemark{\star}   & 0.73 &  1199 &  6.2 &  9.34 &   9.95 &  38.77  & M94  &  \cr
NGC 5929      			& 0.10 &  2492 &  6.2 &  9.57 &  10.21 &  39.21  & M94  &  \cr
NGC 6300      			& 0.42 &  1110 &      &  9.20 &   9.80 &         &      &  \cr
NGC 6890      			& 0.17 &  2419 &  4.2 &  9.44 &   9.87 &  38.43  & M94  &  \cr
NGC 7130\tablenotemark{\star}   & 0.12 &  4842 &  6.6 & 10.28 &  11.05 &  39.73  & S98  &  \cr
NGC 7212      			& 0.31 &  7984 &  5.0 & 10.24 &  10.74 &  40.61  & M94  & T95 \cr
NGC 7582\tablenotemark{\star}   & 0.06 &  1575 &  8.3 &  9.74 &  10.54 &  39.11  & M94  &  \cr
\enddata
\tablenotetext{\star}{Confirmed starburst/Seyfert 2 composite}
\tablecomments{(1) Object name; (2) Galactic B band extinction (from
NED); (3) $cz$ in km$\,$s$^{-1}$; (4) Balmer decrement (compiled from the
literature); (5) $\nu L_\nu$ at 12$\mu$m from IRAS, in L$_\odot$; (6) far
IR luminosity in L$_\odot$ computed with the IRAS 60 and 100 $\mu$m fluxes
using the formula in Sanders \& Mirabel (1996). When only upper limits were
available for the IRAS fluxes we use half the quoted limit to compute
fluxes. (7) H$\beta$ luminosity in erg$\,$s$^{-1}$, corrected only for
Galactic extinction; (8) reference for the Balmer decrement; (9) references
regarding spectropolarimetry detection of hidden Seyfert 1 nucleus.  Code
for references in columns (8) and (9): AM85- Antonucci \& Miller (1985);
B99- Bassani \etal (1999); D88- Dahari \& De Robertis (1988); dG92 de Grijp
\etal (1992); H97- Heisler, Lumsden \& Bailey (1997); K94- Kay (1994); M00-
Moran \etal (2000); M94- Mulchaey \etal (1994); MG90- Miller \& Goodrich
(1990); S98- Schmitt (1998, PhD Thesis); T95- Tran (1995a,b).
}
\label{tab:TABELAO2}
\end{deluxetable}

\subsection{Composite and ``pure'' Seyfert 2's}

\label{sec:CompositesAndPureSeyfert2s}

Fifteen of our Seyfert 2's have unambiguous evidence for starbursts as
revealed by at least one of the 3 diagnostics outlined above.  These are
ESO 362-G8, IC 3639, Mrk 1, Mrk 78, Mrk 273, Mrk 477, Mrk 463E, Mrk 533,
Mrk 1066, Mrk 1073, Mrk 1210, NGC 5135, NGC 5643, NGC 7130 and NGC 7582
(observed prior to the type transition reported in Aretxaga \etal 1999),
which will henceforth be tagged {\it composite} systems. The circum-nuclear
starbursts in these galaxies cover a wide range of properties, from very
young bursts still in the WR phase (e.g., Mrk 477, Mrk 1) to more mature
systems in a ``post-starburst'' phase with deep high order Balmer
absorption lines (e.g., Mrk 78, ESO 362-G8), and systems which have a
mixture of both young and intermediate age stars (e.g., NGC 5135, NGC
7130).

The remaining 20 sources will be labeled ``{\it pure}'' Seyfert
2's. This denomination is used simply to indicate that no clear signs
of starburst activity have been detected in our previous studies. The
spectra among our ``pure'' Seyfert 2's vary from galaxies whose
optical light is entirely dominated by an old, red stellar population
(e.g., NGC 1358 and NGC 5929) to galaxies which show clear signs of an
UV-excess (e.g., Mrk 3 and Mrk 348). It is nearly impossible to tell
on the basis of optical spectra alone whether this FC is associated
with scattered light, a starburst or a combination of both. This point
was discussed at length by SB00, who argue that as many as 30\% of
Seyfert 2's can fall in this ambiguous category. GD01 also finds that
several galaxies are equally well described in terms of a scattered
power-law or a starburst component, depending basically on what is
chosen to represent the underlying old stellar population.  Far-UV
spectroscopy offers the best hope to break this degeneracy, but this
is not practical with currently available instrumentation, so we will
have to live with this ambiguity for some time. Indeed, this ambiguity
pervades all discussions in this paper.

We emphasize that the composite and ``pure'' categories are {\it not}
meant as new elements in the AGN taxonomy. It is clear that this
classification reflects a {\it detectability} effect, since we define
composites as systems in which stars with ages of $10^{6-8}$ yr have been
conclusively detected. This definition therefore introduces an
inevitable bias towards starbursts that are powerful in contrast with the other
components of Seyfert 2 spectra, a bias which will become evident in
several of the results presented in this paper.  While it is likely
that some ``pure'' sources eventually be ``up-graded'' to composites
with deeper or different observations (e.g., when far-UV spectra are
obtained, or when near-IR mass to light ratios are measured), it is
unlikely that any of the composites will ever be reclassified as a
``pure'' Seyfert 2. Until such further notice, all galaxies not in the
list of 15 composites above will be designated ``pure'' Seyfert 2's.

The composite/``pure'' classification will be used throughout this
paper as a guide to interpret our results, even though we shall {\it
not} make explicit use of any of the properties on which this
classification is based.  Instead, we seek to find where composite
sources fit into more general and easily accessible properties of
Seyfert 2's.

\subsection{Comparison samples}

\label{sec:ComparisonSamples}

\begin{figure}
\plotone{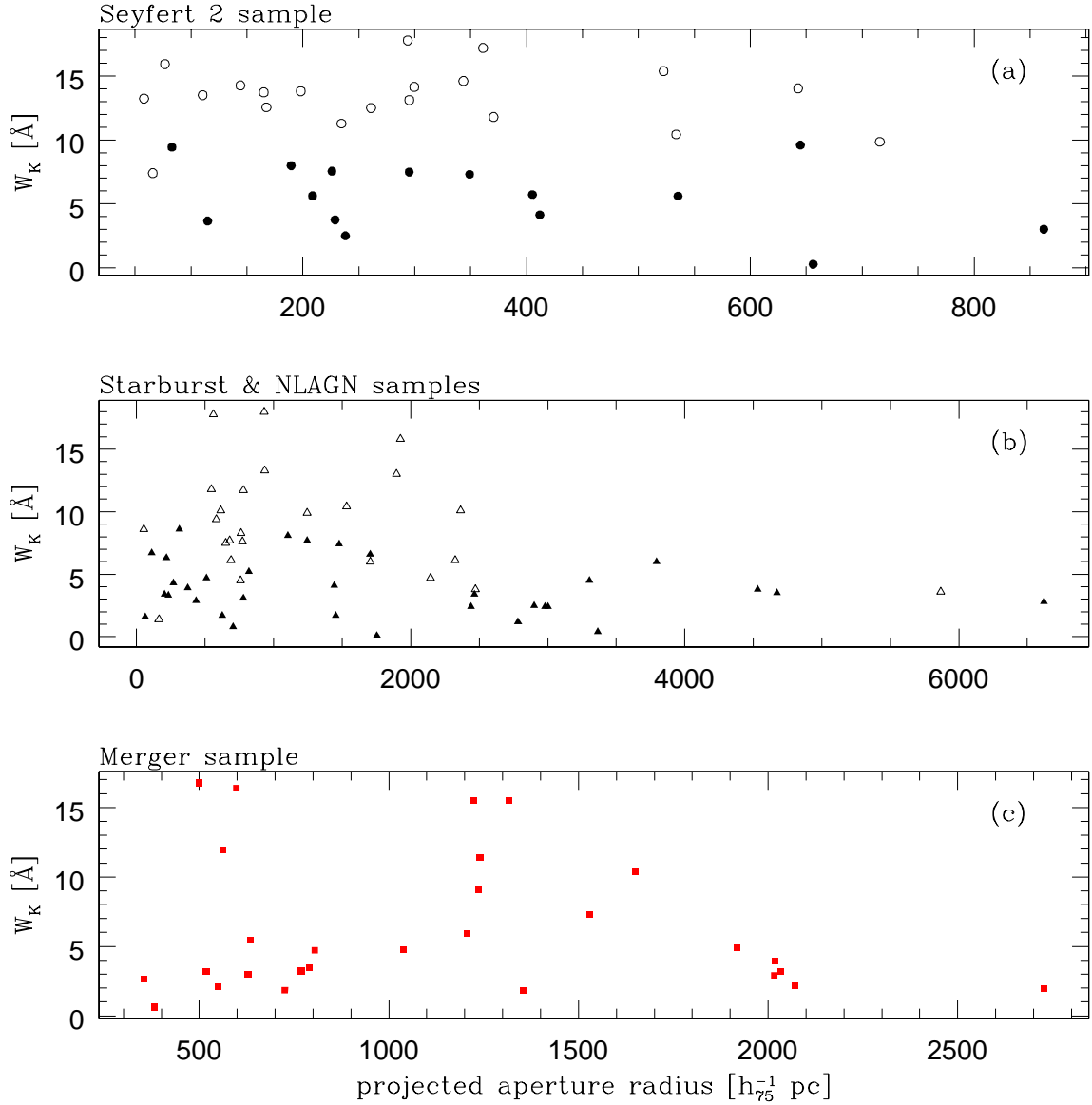}
\caption{Aperture radii against the equivalent width of CaII K for the
Seyfert 2 (a), Starburst and NLAGN (b) and Merger (c) samples.  Empty and
filled circles in panel (a) indicate ``pure'' and composite Seyfert 2's
respectively. Empty triangles in (b) correspond to NLAGN sources, whereas
filled triangles correspond to Starburst galaxies.}
\label{fig:Aperture_X_WK}
\end{figure}

As comparison samples we draw data from three different observational
studies. The first is the spectroscopy of 61 Starburst and Narrow Line
Active Galaxies (NLAGN) by Storchi-Bergmann, Kinney \& Challis (1995) and
McQuade, Kinney \& Calzetti (1995, hereafter collectively referred to as
SBMQ95). The Starburst galaxies in their study will be referred to as the
``Starburst sample'', which will be extensively used as a benchmark when
gauging the results for our Seyfert 2's.  UGCA410 was removed from our
analysis because of apparently incompatible values of $W_K$ and $W_G$,
leaving a total of 36 Starburst galaxies. The remaining 24 sources in
SBMQ95 contain a mixture of Seyfert 2's, LINERs and starburst/AGN
composites, so we give them a generic denomination of ``NLAGN sample''.  In
principle these galaxies serve as a test-bed to evaluate the efficacy of
the diagnostics of compositeness developed from our Seyfert 2 sample. In
practice, we will make limited use of this sample because of the mixed
nature of its sources, which complicates the interpretation.  Galaxies in
the Starburst and NLAGN samples were observed with very large
$10^{\prime\prime} \times 20^{\prime\prime}$ apertures to match the IUE
observations, sampling physical scales of typically $\sim$1 kpc in radius
at the distances of the sources.

A third comparison sample used here is the ``Merger sample'' from the
study of Liu \& Kennicutt (1995a,b, hereafter LK95).  This sample
comprises 28 nuclear optical spectra integrated over a region
corresponding to $\sim 0.5$--2 kpc in radius.  The emission line
properties of these galaxies are quite heterogeneous, comprising 5
Seyfert 2's, 9 HII nuclei and 14 LINERs. Lutz, Veilleux \& Genzel
(1999) have used ISO data to show that IR-luminous galaxies with HII
or LINER classifications are predominantly powered by starbursts, while
dusty AGN are energetically significant in those classified as Seyfert
2's. Thus, we will use the Merger sample to empirically define the
properties of powerful merger-induced dusty starbursts.

The data which will be used from these samples consists of fluxes and $W$'s
of [OII], H$\beta$ and [OIII]; $W_K$, $W_{CN}$ and $W_G$, plus the
continuum fluxes at 3660, 4020, and 4510 \AA\ and IRAS fluxes. Both the
SBMQ95 and LK95 data offer more information than this, but we shall limit
our analysis to the properties above for consistency with the Seyfert 2
sample.

Fig.~\ref{fig:Aperture_X_WK} shows the projected aperture radii, defined as
$r_{ap} = ($slit area$/\pi)^{1/2}$, against $W_K$ for the Seyfert 2, NLAGN,
Starburst and Merger samples. Though the physical region sampled in the
comparison samples extend well beyond that covered by the nuclear spectra
of the Seyfert 2 sample, there is some overlap at small $r_{ap}$ which
guarantees the fairness of the comparisons presented in this paper, and in
any case aperture effects will be discussed whenever relevant.

Fig.~\ref{fig:Aperture_X_WK}a also shows that the composite
classification is {\it not} related to the physical size of the region
sampled by our spectra.  In principle one would expect that systems
observed through larger apertures are more prone to exihibit starburst
features due to the inclusion of light from off-nuclear star-forming
regions. This effect has been quantified by Heckman \etal (1995) in
their analysis of the aperture dependent properties of NGC 1068, whose
1 kpc ring of HII regions dominates the light for observations taken
through kpc-scale apertures. Such rings are in fact rather common
(e.g., Pogge 1989; Wilson \etal 1991; Storchi-Bergmann, Wilson \&
Baldwin 1996; Colina \etal 1997), and thus can play a role in defining
a composite nature for more distant objects, as also discussed by
Heckman etal (1995). If such aperture effects were pronounced in the
Seyfert 2 sample we would expect composites to be concentrated towards
large $r_{ap}$, whereas Fig.~\ref{fig:Aperture_X_WK}a shows that
composites and ``pure'' Seyfert 2's are similarly distributed as a
function of $r_{ap}$. Furthermore, most of the apertures for the
Seyfert 2 sample are much smaller than the 1 kpc typical of
star-forming rings. This effect does not affect the Seyfert 2 sample,
as our circum-nuclear starbursts are relatively compact ($\sim$ few
hundred pc).  These dimensions have in fact been confirmed by HST UV
imaging for Mrk 477 (Heckman \etal 1997), NGC 5135, NGC 7130 and IC
3639 (Gonz\'alez Delgado \etal 1998), and are comparable to the inner
star-forming structures identified in the Circinus galaxy (Wilson
\etal 2000).  A further example of how compact circum-nuclear
starbursts around AGN can be is given in Shields \etal (2000).

Another dataset that will be used in this paper is that of Jansen
\etal (2000). This sample contains multicolor surface photometry for
198 galaxies, most of which classified as normal. We used their data to
calculate the 3660 \AA\ surface brightness of these galaxies inside radii
of 250 and 500 pc, similar to the apertures used for our galaxies.  These
values are used to compare the UV surface brightness of Seyfert and normal
galaxies. The 4 Seyfert 1 and 1 BL Lac in their sample were excluded from
our analysis.

\section{Analysis of the Starlight and the Featureless Continuum
in Seyfert 2's}

\label{sec:Synthesis}

Determining the FC contribution to the optical spectrum of Seyfert 2's
is an important but difficult task, for which there is no definitive
recipe. The traditional and still most widely employed method, which
dates back to at least Koski (1978), is to remove the contribution of
the old red stars from the host's bulge by adopting an elliptical or
early type spiral normal galaxy as a spectral template. The spectral
decomposition is accomplished by assuming a $F_\nu \propto
\nu^{-\alpha}$ power-law for the ``non-stellar'' FC, whose strength is
then adjusted to minimize the residuals, yielding typically $\alpha =
1$--2.  Naturally, this procedure precludes the detection of young
stellar populations, whose most conspicuous optical absorption
features are masked by emission lines even in Starburst
galaxies. Furthermore, there is no simple way to tell apart a power
law FC from that produced by a reddened starburst (e.g., Cid Fernandes
\& Terlevich 1995).  Only upon careful scrutiny of high quality
spectra is it possible to discern the presence of a starburst from its
high order Balmer lines or, sometimes, the WR bump (see discussion in
GD01). Therefore, {\it the FC strength will in general include a
contribution from young stars as well as a \it bona fide non-stellar
FC}, composed of scattered light from the hidden nucleus plus nebular
continuum.

``Featureless'' is of course a misnomer for a spectral component at
least partly associated with starlight.  We nonetheless retain this
terminology both for historical reasons and because to first order a
young stellar population does produce weak spectral features in the
optical range. When referring to scattered light from the hidden
nucleus we will use the term `FC1' introduced by Tran (1995c). As
argued by Cid Fernandes \& Terlevich (1995), Heckman \etal (1995) and
Tran (1995c), strong FC Seyfert 2's must also contain another source
of FC photons, `FC2' in Tran's notation, to account for the absence of
conspicuous broad lines and the larger polarizations observed in the
reflected broad lines than in the reflected nuclear continuum. At
least for the composite systems, there is little doubt that FC2 is
associated with circum-nuclear starbursts (see GD01). Nebular
continuum and emission from the scattering region itself (Tran 1995c)
are other possible contributors to FC2.

In this paper we explore a different way to determine the FC strength. We
decompose the total light into a base of 12 simple stellar population
components represented by star clusters spanning the $10^6$ to $10^{10}$ yr
age range and different metallicities (Schmidt \etal 1991; Bica 1988), plus
a $\nu^{-1.5}$ power-law to represent FC1 and any other AGN component. This
is done by means of a population synthesis analysis similar to that
employed by Schmitt \etal (1999), but performed with the probabilistic
formalism described in Cid Fernandes
\etal (2001), extended to include the power-law component.  Note that
this method is analogous, but not technically equivalent to a full
spectral decomposition as only a few $W$'s and colors are actually
synthesized. As input data we use the $W$'s of CaII~K, CN and G-band,
plus the $F_{3660} /F_{4020}$ and $F_{4510} / F_{4020}$ colors. The
errors on these quantities were set to 0.5 \AA\ for $W_K$ and $W_G$, 1
\AA\ for $W_{CN}$ and 0.05 for the colors; these errors are consistent
with the quality of the spectra, and in any case they do not affect
our conclusions. The output is a 13-D population vector ${\bf x}$
which contains the expected values of the fractional contribution of
each component to the total light at a reference normalization
wavelength. In what follows we shall use 4861 \AA\ as our reference
wavelength unless noted otherwise. Uncertainties in ${\bf x}$ are also
provided by the code.

It is clear from the outset that with so little input information we
will not be able to recover all 13 components of ${\bf x}$ accurately,
but, as discussed by Cid Fernandes \etal (2001), we may rightfully
hope to achieve a coarse yet useful description of the population
mixture by grouping similar elements of the population vector.  In
what follows we shall make use of only three grouped ${\bf x}$
components: $x_{OLD}$, made up from the sum of all base components
with age $\ge 1$ Gyr, $x_{INT}$, corresponding to the
``post-starburst'' $10^8$ yr intermediate age bin, and, most
importantly for our purposes, $x_{FC}$, containing the total
contribution of $\le 10^7$ yr stars plus the power-law component.
These quantities are linked by the $x_{OLD} + x_{INT} + x_{FC} = 1$
normalization constraint, so in practice our method characterizes the
spectral mixture in terms of {\it 2 parameters}.  This bi-parametric
description of the data is in many ways analogous to a Principal
Component Analysis (Sodr\'e \& Cuevas 1997; Rodrigues-Lacerda 2001),
except that by construction we have {\it a priori} knowledge of the
physical meaning of our components.

Note that {\it we do not attempt to disentangle the starburst
and power-law components} explicitly; these are merged in $x_{FC} =
x_{YS} + x_{PL}$, where $x_{YS}$ and $x_{PL}$ stand for the light
fractions due to young ($\le 10^7$ yr) stars and the power-law
respectively. As shown below, both the FC strength and the emission
line properties give us strong hints as to which of these two
components dominates $x_{FC}$, thus helping to break this degeneracy.

\subsection{Results of the Synthesis}

\begin{deluxetable}{lrrr}
\tabletypesize{\scriptsize}
\tablecaption{Synthesis Results}
\tablewidth{0pc}
\tablehead{
\colhead{Galaxy}&
\colhead{$x_{OLD}$ [\%]}&
\colhead{$x_{INT}$ [\%]}&
\colhead{$x_{FC}$ [\%]}\cr
\colhead{(1)}&
\colhead{(2)}&
\colhead{(3)}&
\colhead{(4)}}
\startdata
Mrk 477\tablenotemark{\star}            & 10.5 $\pm$ 3.4 &  2.8 $\pm$ 2.0 & 86.6 $\pm$ 4.2  \cr
Mrk 463E\tablenotemark{\star}            & 27.7 $\pm$ 4.3 &  3.8 $\pm$ 2.5 & 68.5 $\pm$ 4.0  \cr
NGC 5135\tablenotemark{\star}            & 27.2 $\pm$ 7.3 & 26.9 $\pm$ 6.2 & 45.9 $\pm$ 5.8  \cr
NGC 7130\tablenotemark{\star}            & 36.3 $\pm$ 8.0 & 23.6 $\pm$ 6.1 & 40.1 $\pm$ 5.9  \cr
Mrk 533\tablenotemark{\star}            & 40.9 $\pm$ 7.7 & 25.3 $\pm$ 6.1 & 33.8 $\pm$ 5.5  \cr
Mrk 1066\tablenotemark{\star}           & 51.5 $\pm$ 6.8 & 16.3 $\pm$ 5.4 & 32.3 $\pm$ 4.9  \cr
NGC 7582\tablenotemark{\star}             & 35.6 $\pm$ 7.8 & 33.8 $\pm$ 6.5 & 30.6 $\pm$ 5.7  \cr
Mrk 1\tablenotemark{\star}              & 63.1 $\pm$ 5.6 &  7.1 $\pm$ 3.6 & 29.8 $\pm$ 4.0  \cr
Mrk 1073\tablenotemark{\star}           & 55.3 $\pm$ 6.2 & 15.6 $\pm$ 5.1 & 29.1 $\pm$ 4.6  \cr
Mrk 1210\tablenotemark{\star}            & 65.1 $\pm$ 4.7 &  5.9 $\pm$ 3.1 & 29.0 $\pm$ 3.5  \cr
NGC 1068             & 63.2 $\pm$ 5.5 & 11.1 $\pm$ 4.3 & 25.7 $\pm$ 4.0  \cr
Mrk 3                & 77.8 $\pm$ 2.7 &  1.1 $\pm$ 0.8 & 21.1 $\pm$ 2.0  \cr
NGC 7212             & 77.3 $\pm$ 3.6 &  2.0 $\pm$ 1.4 & 20.8 $\pm$ 2.6  \cr
Mrk 273\tablenotemark{\star}            & 47.3 $\pm$ 7.1 & 34.8 $\pm$ 6.2 & 17.9 $\pm$ 4.7  \cr
Mrk 34               & 78.1 $\pm$ 4.6 &  4.1 $\pm$ 2.5 & 17.7 $\pm$ 3.1  \cr
IC 3639\tablenotemark{\star}            & 69.9 $\pm$ 6.8 & 14.4 $\pm$ 5.0 & 15.7 $\pm$ 4.1  \cr
Mrk 348              & 87.2 $\pm$ 4.4 &  3.2 $\pm$ 2.0 &  9.7 $\pm$ 2.6  \cr
NGC 2110             & 87.8 $\pm$ 3.4 &  2.5 $\pm$ 1.6 &  9.6 $\pm$ 2.0  \cr
IC 1816              & 84.5 $\pm$ 4.8 &  7.2 $\pm$ 3.3 &  8.3 $\pm$ 2.6  \cr
NGC 5643\tablenotemark{\star}             & 73.2 $\pm$ 7.2 & 19.5 $\pm$ 5.3 &  7.3 $\pm$ 3.2  \cr
Mrk 78\tablenotemark{\star}               & 72.0 $\pm$ 6.5 & 21.2 $\pm$ 4.9 &  6.8 $\pm$ 2.9  \cr
NGC 3081             & 90.1 $\pm$ 4.3 &  3.4 $\pm$ 2.0 &  6.4 $\pm$ 2.1  \cr
Mrk 573              & 90.9 $\pm$ 4.3 &  3.4 $\pm$ 2.0 &  5.6 $\pm$ 1.9  \cr
NGC 5929             & 91.7 $\pm$ 3.7 &  2.7 $\pm$ 1.7 &  5.6 $\pm$ 1.9  \cr
NGC 6890             & 93.1 $\pm$ 4.5 &  1.9 $\pm$ 1.4 &  5.0 $\pm$ 2.2  \cr
NGC 1386             & 89.9 $\pm$ 5.4 &  5.1 $\pm$ 2.7 &  5.0 $\pm$ 2.1  \cr
Mrk 607              & 90.4 $\pm$ 4.4 &  4.7 $\pm$ 2.4 &  4.9 $\pm$ 1.8  \cr
ESO 362-G8\tablenotemark{\star}           & 56.6 $\pm$ 5.9 & 38.5 $\pm$ 4.9 &  4.9 $\pm$ 2.4  \cr
CGCG 420-015         & 91.7 $\pm$ 5.0 &  4.1 $\pm$ 2.3 &  4.2 $\pm$ 1.8  \cr
MCG -05-27-013       & 93.9 $\pm$ 3.7 &  1.9 $\pm$ 1.3 &  4.2 $\pm$ 1.6  \cr
IRAS 11215-2806      & 90.0 $\pm$ 5.7 &  5.9 $\pm$ 3.0 &  4.0 $\pm$ 1.9  \cr
ESO 417-G6           & 92.7 $\pm$ 4.6 &  4.0 $\pm$ 2.2 &  3.3 $\pm$ 1.5  \cr
Fairall 316          & 96.6 $\pm$ 3.8 &  1.5 $\pm$ 1.0 &  1.9 $\pm$ 0.9  \cr
NGC 6300             & 96.5 $\pm$ 4.6 &  1.7 $\pm$ 1.2 &  1.8 $\pm$ 1.0  \cr
NGC 1358             & 97.6 $\pm$ 3.8 &  1.2 $\pm$ 0.8 &  1.2 $\pm$ 0.7  \cr
\enddata
\tablenotetext{\star}{Confirmed Starburst/Seyfert 2 composites}
\tablecomments{(1) Object name; (2--4) fractions of the total light at
4861 \AA\ due to old stars ($x_{OLD}$), 100 Myr stars ($x_{INT}$); and
$\le 10$ Myr old stars plus a $\nu^{-1.5}$ power-law FC ($x_{FC}$).
Galaxies are organized in order of descending FC strength.}
\label{tab:Synthesis}
\end{deluxetable}

\begin{figure}
\epsscale{0.95}
\plotone{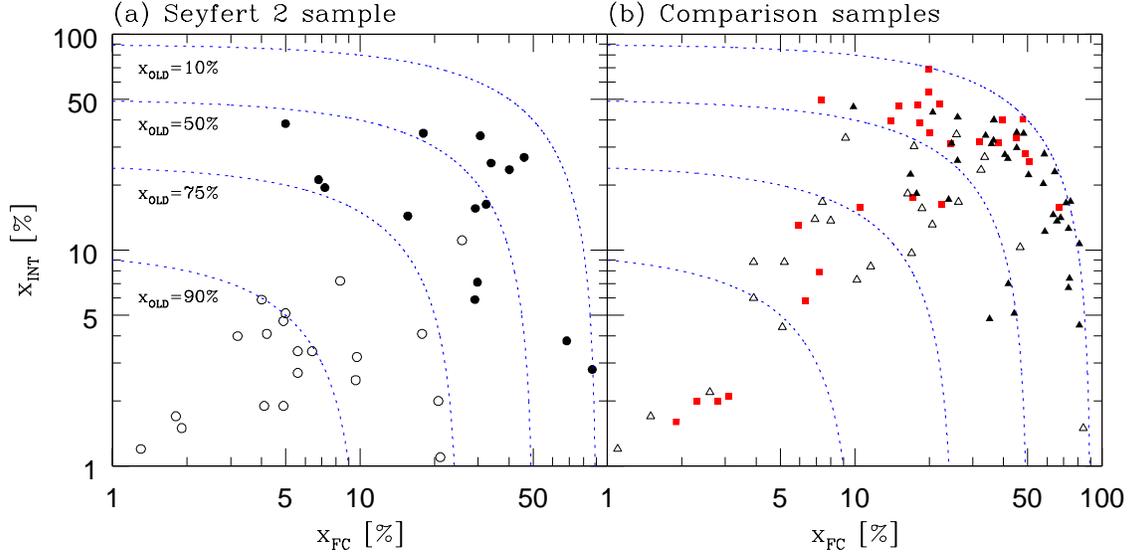}
\caption{Results of the synthesis analysis, condensed into a
bi-parametric $x_{FC}$ and $x_{INT}$ representation. Dotted lines trace
lines of constant $x_{OLD}$, as labeled. (a) Sources from the Seyfert 2
sample, with filled circles indicating the starburst/Seyfert 2
composites. The empty circle among the $x_{OLD} < 75\%$ zone otherwise
exclusively occupied by composites is NGC 1068. As $x_{OLD}$ increases it
becomes progressively difficult to detect starburst features in Seyfert
2's, explaining the clear separation between composites and ``pure''
Seyfert 2's in this diagram. (b) Results for the Starburst (filled
triangles), NLAGN (empty triangles) and Merger (filled squares) samples.}
\label{fig:Fig_xFC_X_xINT}
\end{figure}

The results of this analysis are listed in Table~\ref{tab:Synthesis}
in order of decreasing $x_{FC}$.  A first noticeable result in this
Table is that composites are concentrated towards the top of the list:
All 10 $x_{FC} \ge 29\%$ sources are confirmed Seyfert 2/starburst
composites, and 12 of the 15 composites are among the 16 sources with
$x_{FC} > 15\%$ at 4861 \AA.  The identification of $x_{FC}$ with
$x_{YS}$ (i.e., $x_{YS} \gg x_{PL}$) is thus safe for these objects.

An even more remarkable result is that {\it all} 15 composites but {\it
only one} ``pure'' Seyfert 2 (NGC 1068) have $x_{OLD} < 75\%$.  This is
illustrated in Fig.~\ref{fig:Fig_xFC_X_xINT}a, where we condense the
results of the synthesis in the $x_{FC} \times x_{INT}$ plane, with dotted
lines indicating lines of constant $x_{OLD} = 1 - x_{INT} - x_{FC}$. {\it
The diagram segregates composite from ``pure'' systems very effectively},
and it does so in agreement with what one would expect: composites have
larger proportions of intermediate age and young stars, the latter being
included in $x_{FC}$.

As a check on our method, we have applied it to the spectra in the
comparisons samples, which, for consistency, were processed in exactly the
same way as our Seyfert 2's.  If our method works, we would expect {\it
bona fide} Starburst galaxies to be located within the $x_{OLD} < 75\%$
zone, probably more towards even lower values of $x_{OLD}$, overlapping
with our more extreme composites. This is confirmed in
Fig.~\ref{fig:Fig_xFC_X_xINT}b, which shows the synthesis results for the
Starburst, Merger and NLAGN samples.  There is not a single Starburst
galaxy (filled triangles) for which $x_{OLD} > 65\%$, whereas most of the
NLAGN are located there. Furthermore, most of the NLAGN which intrude into
the Starburst region are known starburst/Seyfert 2 composites!  The four
empty triangles in the $x_{OLD} < 50\%$ zone, for instance, are Mrk 477,
NGC 5135, NGC 7130 (all also in our Seyfert 2 sample) and NGC 7496, whose
composite nature is discussed in V\'eron, Gon\c{c}alves \& V\'eron-Cetty
(1997). The $50\% < x_{OLD} < 75\%$ zone also contains known composites,
both from our sample (IC 3639 and NGC 7582), and from independent work,
such as NGC 4569 (Keel 1996; Maoz \etal 1998), NGC 1672 (V\'eron \etal
1997) and NGC 6221 (Levenson \etal 2001b).

As expected, most of the sources in the Merger sample (squares in
Fig.~\ref{fig:Fig_xFC_X_xINT}b) behave like starbursts in our stellar
population diagram. Note that our synthesis indicates a strong
contribution of $\sim 10^8$ yr populations for most sources in this
sample.  This is consistent with the results of LK95, who find a high
incidence of systems in a post-starburst phase based on the detection
of the Balmer absorption series. Those mergers which are dominated by
old stars ($x_{OLD} > 50\%$) are predominantly LINERs (e.g., NGC 942,
NGC 3656 and 3C293).

It is thus clear that, despite its limitations, our bi-parametric
description of the data provides a very efficient empirical diagnostic
of compositeness in Seyfert 2's.  Since our operational definition of
composites is based on the detectability of starburst features, which
is certainly facilitated when they make a strong contribution to the
continuum, the result that sources with large $x_{FC}$ and/or
$x_{INT}$ are mostly composites is somewhat redundant.  The importance
of this result is twofold. First, it shows a striking empirical
similarity between the {\it bona fide} Starburst galaxies (whose
optical/near-UV continuum is definitely produced by young and
intermediate-age stars) and the starburst/Seyfert 2 composites.
Second, it demonstrates an excellent {\it consistency} between the two
techniques, which are based on different observables. The synthesis
process did {\it not} use any of the information used in the original
identification of starbursts in these composites, namely, far-UV
stellar wind features, high order Balmer lines, and/or the WR bump,
all of which are much more subtle or hard to obtain than the
observables used as input for the synthesis code.

\subsection{Contrast and Evolution}

\label{sec:Constrast_and_Evolution}

The separation between composites and ``pure'' Seyfert 2's in
Fig.~\ref{fig:Fig_xFC_X_xINT}a strongly suggests that, as anticipated in
\S\ref{sec:CompositesAndPureSeyfert2s}, the composite/``pure''
classification is mostly driven by a {\it contrast} effect: Systems with
circum-nuclear starbursts residing in galaxies where the old stellar
population dominates the spectrum would simply not be recognized as
composite systems. This contrast effect is nicely quantified by the value
of $x_{OLD}$, which can therefore be read as a measure of the difficulty to
detect any non-trivial spectral component and to interpret its origin.
Fig.~\ref{fig:Fig_xFC_X_xINT} shows that our current techniques establish a
threshold of $x_{OLD} \sim 75\%$, beyond which we are not able to recognize
circum-nuclear starbursts. For instance, if we scale down the
circum-nuclear starburst in IC 3639 by a factor $\ge 2$, it would move from
$x_{OLD} = 70\%$ to more than 82\%, where, judging by
Fig.~\ref{fig:Fig_xFC_X_xINT}, we would classify it as a ``pure'' Seyfert
2.  It is therefore likely that at least some ``pure'' Seyfert 2's are just
that: composites with weak starbursts.

The presence of strong HI emission lines introduces another contrast
effect, since they dilute the high order Balmer absorption lines, thus
hindering the use of this diagnostic of starbursts. Therefore {\it the
combination of strong emission lines and a large $x_{OLD}$ is highly
unfavorable to the detection of circum-nuclear-starbursts}.  We will
return to this issue in \S\ref{sec:Difficulty_to_detect_starbursts},
after we analyze the emission line data.

Besides the contrast between the starburst and the underlying old
stellar population, {\it evolution} is the main property defining the
location of a source in Fig.~\ref{fig:Fig_xFC_X_xINT}. One can easily
imagine an evolutionary sequence which starts at some large value of
$x_{FC}$ in the bottom right part of the diagram, with massive young
starts dominating the spectrum and no intermediate age population, and
gradually moves towards large $x_{INT}/x_{FC}$ (i.e., to the top left
of the diagram) on $\sim 10^8$ yr time scales. The exact shape of
evolutionary tracks in our $x_{FC} \times x_{INT}$ diagram is dictated
by the detailed star-formation history, but for the purposes of the
qualitative discussion below one can imagine that such tracks broadly
follow the dotted lines traced in Fig.~\ref{fig:Fig_xFC_X_xINT} up to
a few hundred Myr, and then collapse to the origin, after the
starburst ends and its stars move into our old population bin.

Here too we find a remarkable agreement between the present synthesis
analysis and the results of our previous work on the composites in the
Seyfert 2 sample.  The four composites with larger $x_{FC} / x_{INT}$
are Mrk 477, Mrk 463E, Mrk 1210 and Mrk 1 (going upwards in
Fig.~\ref{fig:Fig_xFC_X_xINT}a). These are precisely the 4 systems
where we have identified WR stars (Heckman \etal 1997;
Storchi-Bergmann \etal 1998; GD01), signposts of very recent
star-formation. They therefore rank as the youngest circum-nuclear
starbursts among our composites, in agreement with their location in
Fig.~\ref{fig:Fig_xFC_X_xINT}a. At the other extreme, the three
composites with $x_{FC} < 10\%$ are NGC 5643, Mrk 78 and ESO 362-G8,
all of which present spectroscopic signatures of a dominant
`post-starburst' population (Schmitt \etal 1999; SB00; GD01), and rank
as our oldest composites, in agreement with the large $x_{INT}$ but
small $x_{FC}$ we obtain.  Given this agreement at the young and old
ends of a starburst evolutionary sequence, we would expect that
systems located at $x_{FC} / x_{INT}$ values in between these extremes
present a mixture of young and intermediate age stars. This is indeed
the case. The cluster of 4 composites towards the top right in
Fig.~\ref{fig:Fig_xFC_X_xINT}a, for instance, contains NGC 5135, NGC
7130, NGC 7582 and Mrk 533, all of which reveal spectroscopic
signatures of this mixture (SB00; GD01), such as pronounced high order
Balmer absorption lines simultaneous with far-UV stellar wind lines.

What makes this agreement remarkable is the fact that while the
characterization of the starburst component in our earlier work involved a
detailed spectroscopic analysis, the synthesis performed here used just
three $W$'s of metallic absorption bands plus a couple of near-UV colors.
Our quicker and cheaper method is therefore not only able to recognize
composite systems from a handful of easily measured observables, but it
allows us to go one step further to provide a rough description of
evolutionary state of the starburst component.

We will return to the issue of evolution in
\S\ref{sec:Discussion}. Before that, we have to deal with the
contribution of the non-stellar FC in $x_{FC}$.  This issue was
deliberately omitted from the discussion above because we do {\it not}
have a recipe to separate the stellar and non-stellar parts of the FC,
neither do we know how or whether the non-stellar FC evolves on
time-scales comparable to the starburst lifetime.  These difficulties
limit considerations about evolution to composites alone, for which we
know that the starburst component is the dominant contributor to the
FC.

\subsection{The FC component}

Our inability to disentangle the starburst and non-stellar components
of the FC is currently the major obstacle to a full characterization
of the starburst-AGN connection in Seyfert 2's.  Unfortunately, these
are also the most relevant spectral components in Seyfert 2's. The
starburst portion of $x_{FC}$ traces the ongoing star-formation, being
therefore associated with young massive stars which may have a
significant impact on the ionization of the gas in the circum-nuclear
environment. On the other hand, the non-stellar component (which is
presumably mainly scattered light, i.e., FC1) is the only continuum
feature associated with the AGN in these galaxies. For these reasons
we concentrate on the FC component throughout most of the rest of this
paper. Whereas little can be said about the starburst and AGN shares
of $x_{FC}$ in ``pure'' Seyfert 2's, composite systems have their FC
dominated by the starburst (GD01; SB00), and hence provide a useful
guide as to what can be deduced from the FC strength alone.

\subsubsection{Strong FC sources are Seyfert 2/starburst composites}

Table~\ref{tab:Synthesis} shows that strong FC sources ($x_{FC} \ga
30\%$) are likely to be composite systems. Within the context of the
unified model, it is not surprising to find that a starburst component
dominates over FC1 whenever the total FC is strong, since scattered
light cannot exceed a fraction of $\sim 30 \%$, otherwise reflected
broad lines should become easily discernible in the direct spectrum
and the galaxy would no longer be classified as a type 2 Seyfert. For
a typical $W$ of $\sim 100$ \AA\ for broad H$\beta$ in Seyfert 1's
(Goodrich 1989; Binette, Fosbury \& Parker 1993), more than $30\%$
scattered light would imply a broad H$\beta$ stronger than 30 \AA\ in
the direct total spectrum.  Cid Fernandes \& Terlevich (1995) argue
that broad lines should be seen even for smaller $x_{FC1}$, but we
figure 30\% is a reasonable limit in practice because NGC 1068 has
$x_{FC} = 26\%$, and broad lines have not been conclusively detected
there without the hindsight of spectropolarimetry to guide the eye
(Miller \& Goodrich 1990; Tran 1995b; Malkan \& Filippenko 1983).

In support of this conclusion, we note that several cases of strong FC
sources reported in the literature turned out to harbor starbursts not
accounted for in the classical starlight template decomposition. For
instance, Miller \& Goodrich (1990) find $x_{FC} = 100\%$ in Mrk 463E,
while GD01 suggest it actually contains a powerful starburst in the WR
phase. The same can be said about Mrk 477, for which Tran (1995a,b) finds
$x_{FC} = 59\%$ at 5500 \AA, and Mrk 1066, for which Miller \& Goodrich
(1990) find a 72\% FC at 5300 \AA.  Even in Cygnus A, whose blue FC has
remained a mystery for more than a decade (e.g., Goodrich
\& Miller 1989; Tadhunter, Scarrott \& Rolph 1990), there now seems to be
evidence for young stars (Fosbury \etal 1999). Indeed, inspection of the
Keck spectrum in Ogle \etal (1997) even indicates the presence of a
possible WR bump! This is consistent with our conclusion, since estimates
of the FC strength for Cygnus A place it above the 30\% level (Osterbrock
1983).

This high ``success''-rate suggests that one can {\it predict} the
existence of circum-nuclear starbursts based only on the FC strength.
Despite the non uniformity of FC estimation methods, one can be fairly
confident that a search for sources with $x_{FC} > 30 \%$ reported in
the literature should identify other composite systems. Kay (1994),
for instance, reports the following Seyfert 2's with an FC stronger
than 30\% at 4400 \AA: Mrk 266SW (30\%), Mrk 1388 (45\%), NGC 591
(46\%), NGC 1410 (38\%), NGC 1685 (38\%), NGC 4922B (53\%), NGC 7319
(35\%), NGC 7682 (30\%). Koski (1978) finds a 34\% FC at 4861 \AA\ for
3C184.1 and 35\% for Mrk 198.  Evidence for young and intermediate age
stars in the spectra of Mrk266SW was reported by Wang \etal (1997).
According to Gon\c{c}alves, V\'eron-Cetty \& V\'eron (1999), this
galaxy also has a composite emission line spectrum, with emission line
ratios intermediate between those of Seyferts and Starbursts. The same
description is given for the interacting galaxy NGC 4922B by
Alonso-Herrero \etal (1999). We could not find information regarding
the compositeness or otherwise of the remaining galaxies. Our
prediction is that most of them should present detectable signatures
of circum-nuclear starburst activity.

\subsubsection{Moderate and weak FC sources are ambiguous}

\label{sec:moderate_and_weak_FC_sources}

While all $x_{FC} \ga 30\%$ sources are composites, {\it the converse is
not true}. For IC 3639, a galaxy whose composite nature has been
conclusively established by both optical and UV spectroscopy, we find a
16\% FC.  The range between 15 and 30\% also contains 4 ``pure'' Seyfert 2
nuclei: NGC 1068, Mrk 3, NGC 7212 and Mrk 34.

NGC 1068 is sometimes listed as an example of a starburst-AGN link
because of its bright ring of HII regions at $\sim 1$ kpc from the
nucleus, which contributes roughly half of the bolometric (IR)
luminosity of this prototype Seyfert 2 (Telesco \etal 1984; Lester
\etal 1987). If observed from further away, the ring would dominate
the optical spectrum, imprinting signatures which would cast distant
NGC 1068's in the composite category (Heckman \etal 1995; Colina \etal
1997), but, as argued in \S\ref{sec:Sample}, this is not the effect
behind the starburst features identified in our composites
(Fig.~\ref{fig:Aperture_X_WK}). The spectrum of NGC 1068 analyzed here
pertains to a much smaller region, corresponding to $r_{ap} = 66$
pc. The only indications of star-formation within this central region
are the strong CaII triplet at $\sim 8500$ \AA\ (Terlevich, D\'{\i}az
\& Terlevich 1990) and the small $M/L$ ratio at 1.6$\mu$m (Oliva \etal
1995, 1999), both of which point to an intermediate-age population,
qualitatively consistent with the $x_{INT} = 11\%$ found in our
synthesis. There does not seem to be substantial ongoing
star-formation associated with this ``post-starburst'' population,
since neither far-UV (Caganof \etal 1991) nor optical (GD01; Miller \&
Antonucci 1983) spectroscopy reveal signs of young stars close to the
nucleus. Furthermore, from the work of Antonucci \& Miller (1985),
Miller \& Goodrich (1990) and Tran (1995c) we know that NGC 1068 does
{\it not} suffer from the ``FC2-syndrome'' (lower polarization in the
continuum than in the broad lines), so most, if not all, of its
nuclear FC is indeed FC1. Since our $x_{FC}$ estimate is in fair
agreement with previous determinations (e.g., the 22 \% at 4600 \AA\
found by Miller \& Antonucci 1983 and the 16\% at 5500 \AA\ derived by
Tran 1995), there is little doubt that $x_{FC1} \gg x_{YS}$ in the
central region of NGC 1068.  We therefore keep NGC 1068 in the
``pure'' Seyfert 2 category.

The nature of the FC is not so clear for the 3 other moderate FC ``pure''
Seyfert 2's in our sample: Mrk 3, NGC 7212 and Mrk 34. In their detailed
analysis of the same data used here, GD01 find that the dilution of the
metal absorption lines in these galaxies is better modeled by an elliptical
galaxy plus power-law spectral decomposition than using an off-nuclear
template, which, together with the lack of starburst features, lead them to
favor a scattered light origin for their FC.  The extended blue continuum
aligned with the radio axis and ionization cone in Mrk 3 (Pogge \& De
Robertis 1993) supports this idea.  Kotilainen \& Ward (1997) find a
similar feature in NGC 7212, though not aligned along the [OIII] emission.
These indications of a FC1 component are confirmed by the
spectropolarimetry observations of Tran (1995a,b,c), which revealed their
hidden Seyfert 1 nuclei.  However, Tran also finds that Mrk 3 and NGC 7212
suffer from acute ``FC2-itus''!  According to his analysis, only 4\% (Mrk
3) and 2\% (NGC 7212) of the flux at 5500 \AA\ is attributable to FC1; the
rest of their FC is `FC2'!  The total FC = FC1 + FC2 fractions at 5500 \AA\
found by Tran are 12\% for Mrk 3 and 17\% for NGC 7212, which are in very
good agreement with our estimates of $x_{FC}$ when translated to the same
wavelength. Therefore, from the point of view of their polarization
spectra, most of their FC still have to be explained.

Below $x_{FC} \sim 30\%$ we therefore enter a ``gray zone'': the
power-law/starburst degeneracy sets in and the identity of the FC
becomes ambiguous. Extra information (spectropolarimetry, UV
spectroscopy, imaging) may help disentangling these two components in
a few sources, but the situation becomes even fuzzier for weak FC
sources ($x_{FC} < 15 \%$). The case of Mrk 348 is illustrative in
this respect.  Using an elliptical galaxy template, Tran (1995a,b)
finds a 27\% FC at 5500 \AA, which, when combined with his
spectropolarimetry data, propagates to the conclusion that 22\% of the
total light is associated with something else, ie., FC2.
Storchi-Bergmann \etal (1998), on the other hand, find a much weaker
FC by using an off-nuclear spectrum as a starlight template, a
conclusion corroborated by GD01 on the basis of independent
observations. In fact, they find that the nuclear spectrum is well
matched with no FC at all, but, guided by the results of Tran, they
favor a model in which 5\% of the light comes from the FC1 component
detected via spectropolarimetry.

These inconsistencies undoubtedly emerge due to the intrinsic weakness of
the FC, which, combined with the dispersion of stellar populations in
active galaxies (Cid Fernandes \etal 1998; Schmitt \etal 1999; Boisson
\etal 2000; GD01), boosts the differences between different FC estimation
methods. Whereas in Mrk 348 the combination of spectropolarimetry and
long-slit spectra allowed considerations about the nature of its FC, we do
not have such information for most sources below Mrk 348 in
Table~\ref{tab:Synthesis}, and even if such data existed, it is clear that
any attempt to perform a detailed spectral decomposition would be highly
uncertain, since it is simply unrealistic to expect accuracies better than
$\sim \pm 5$\% in any $x_{FC}$ determination method. From the point of view
of this paper, we consider all such sources ambiguous.  We hope to shed
light on the nature of these moderate and weak FC Seyfert 2's by
considering information other than that contained in the optical continuum
and absorption lines (\S\ref{sec:Emission_Lines}).

\subsection{Empirical calibration of the FC strength}

\begin{figure}
\plotone{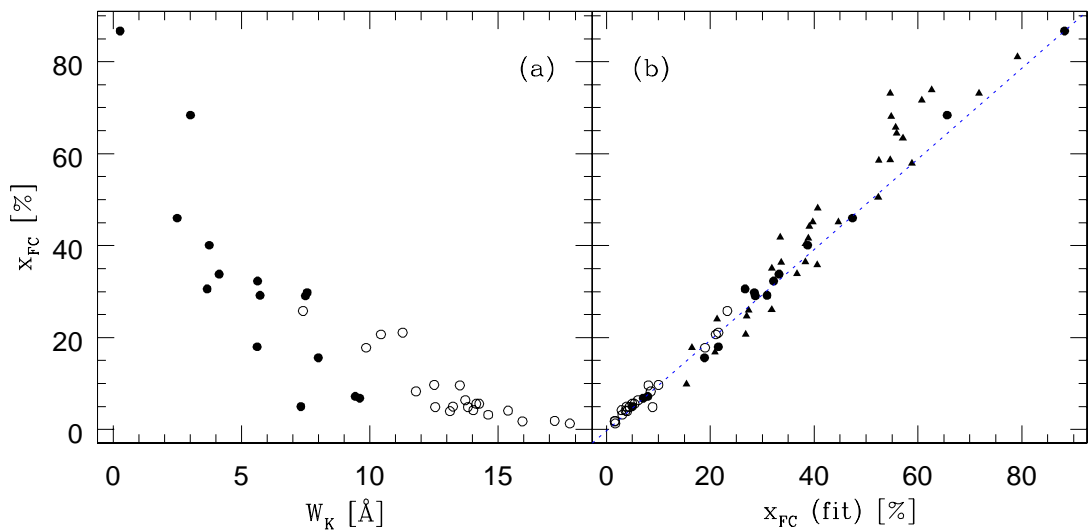}
\caption{(a) Relation between the FC strength at 4861 \AA\ and the
equivalent width of CaII K for the Seyfert 2 sample. Composites are plotted
with filled circles. (b) Comparison of the FC strengths derived from the
synthesis and from the empirical fit of $x_{FC}$ as a function of $W_K$ and
a near UV color (equation~\ref{eq:xFC_calibration}), which offers an easy
way of estimating $x_{FC}$. Starburst galaxies from the SBMQ95 data are
also included in this plot (filled triangles), but not in the fit.}
\label{fig:xFC_callibration}
\end{figure}

As in any other study dealing with optical spectra of Seyfert 2's, the
FC strength plays a major role throughout this paper. The results
above already indicate that, regardless of the ambiguities involved in
the interpretation of $x_{FC}$, our method was able to recover a
meaningful component of the near UV--optical continuum, a conclusion
which will become stronger in the analysis that follows.  This in
itself is an important result, since we based our FC estimation on
just a handful of easily measurable quantities. Though the synthesis
process is rather elaborate, we may use its results to derive an {\it
a posteriori} calibration of $x_{FC}$ in terms of the observables it
is based upon. In fact, as shown in Fig.~\ref{fig:xFC_callibration},
we find that the expression

\begin{equation}
\label{eq:xFC_calibration}
x_{FC}  =
	-0.33 \left( \frac{W_K}{20} \right)
	+ 0.52 \left( \frac{W_K}{20} \right)^2
	+ 0.89 \left( \frac{F_{3660}}{F_{4020}} \right)
	-1.04 \left( \frac{W_K}{20} \right)
     		\left( \frac{F_{3660}}{F_{4020}} \right)
	-0.08,
\end{equation}

\noindent where $W_K$ is in \AA, recovers $x_{FC}$ within $\pm 3\%$ for all
galaxies.  Since the measurements of both $W_K$ and the $F_{3660} / F_{4020}$
color do not require large S/N spectra, this calibration can be applied to
data sets not meeting the S/N $> 25$ standard of our sample (which was
necessary to identify weak starburst features). This provides a much more
straight-forward and well defined way to estimate $x_{FC}$.

Although old stars are of no direct interest in the starburst-AGN
connection, we have seen that they play a central role in defining the
detectability of circum-nuclear starbursts, so it is useful to have an
equation analogous to (\ref{eq:xFC_calibration}) for $x_{OLD}$. We find
that

\begin{equation}
\label{eq:xOLD_calibration}
x_{OLD} =
	1.92 \left( \frac{W_K}{20} \right)
	- 0.98 \left( \frac{W_K}{20} \right)^2
	+ 0.05
\end{equation}

\noindent reproduces the synthetic $x_{OLD}$ to better than $\pm 5\%$ for all
galaxies.

Equations (\ref{eq:xFC_calibration}) and (\ref{eq:xOLD_calibration}) also
do a good job for the Merger, NLAGN and Starburst samples.  This is
illustrated in Fig.~\ref{fig:xFC_callibration}b for the galaxies in the
Starburst sample.  The difference, of course, is that for these {\it bona
fide} Starbursts the FC is undoubtedly produced by young stars.

A corollary of $x_{FC}$ and $x_{OLD}$ being so closely related to $W_K$ is
that this quantity can be used as a first order measure of the stellar
population mix and the spectral dilution of metallic features of old stars
caused by young stars, a power-law or a combination of both. Indeed, in
several of the plots below $W_K$ is used with this function, since, for
most practical purposes, $W_K$ and $x_{FC}$ are equivalent
(Fig.~\ref{fig:xFC_callibration}).

\section{Emission Lines}

\label{sec:Emission_Lines}

Circum-nuclear starbursts in Seyfert 2's act as a second source of
ionizing photons besides the hidden AGN, and as such are bound to
contribute at some level to their emission line spectrum. In this
section we examine our sample in search of signs of this contribution,
analyzing the equivalent widths of strong emission lines
(\S\ref{sec:Ws}), emission line ratios (\S\ref{sec:He2_and_O3}) and
line profiles (\S\ref{sec:LineProfiles}). An investigation of line-FC
relations is also presented (\S\ref{sec:line_FC_relation} and
\S\ref{sec:Line_X_FC_Interpretation}).  When referring to emission
line equivalent widths measured with respect to the FC we will use the
notation $W^{FC}_\lambda \equiv W^{obs}_\lambda / x_{FC}$ to
distinguish them from the observed ones, denoted by $W^{obs}_\lambda$.

\subsection{The Equivalent Widths of H$\beta$, [OIII] and [OII]}

\label{sec:Ws}

The effect of a circum-nuclear starburst on the $W^{FC}_{H\beta}$ of
Seyfert 2's is easily predicted by considering the inverse situation: a
pure starburst to which we add a Seyfert 2. The latter will surely add more
to the $H\beta$ flux than to its underlying continuum, since the
non-stellar FC from the active nucleus is only seen periscopicaly and hence
strongly suppressed in the scattering process, whereas photons from the
Narrow Line Region (NLR) do not suffer such suppression because they are
not obscured by the pc scale dusty torus which surrounds the nucleus. The
net effect of this superposition is that Seyfert 2's should have a larger
ratio of line per FC photons than in pure starbursts.

\begin{figure}
\plotone{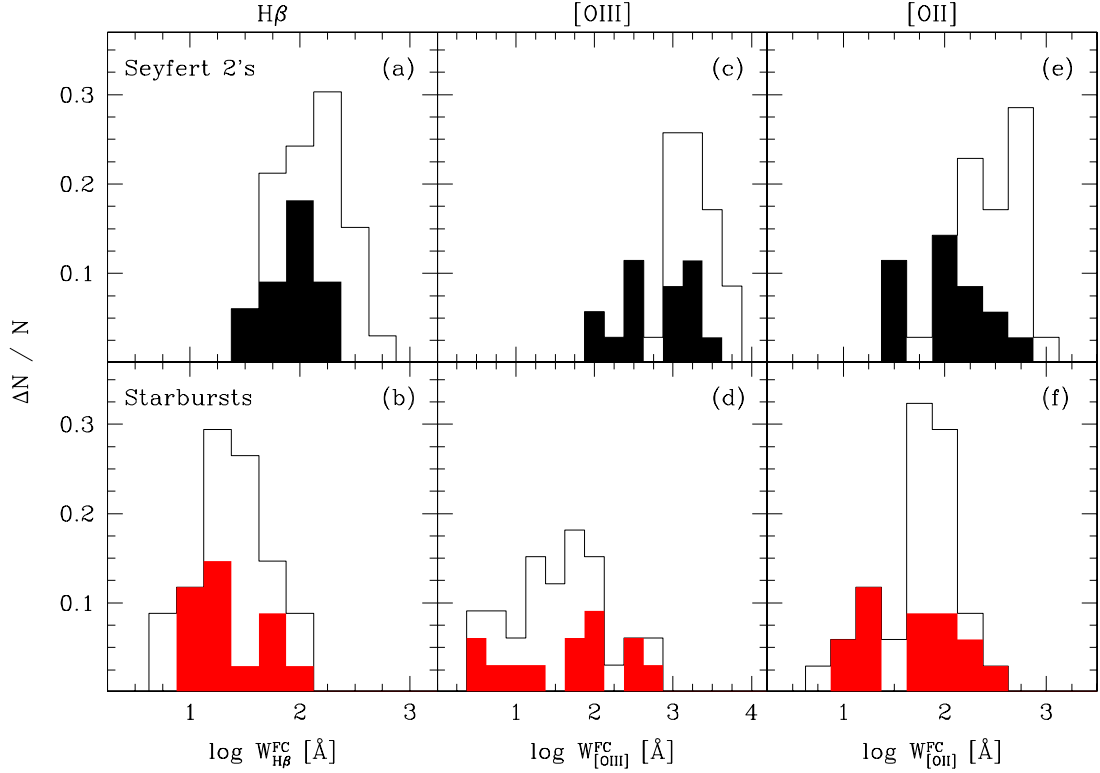}
\caption{{\it Top}: Distribution of the equivalent widths of H$\beta$,
[OIII] and [OII] for the sources in the Seyfert 2 sample. The open
histogram represents all galaxies in the sample, whereas the shaded regions
indicate the composites only. {\it Bottom}: As above for galaxies in the
Starburst sample.  The shaded histogram marks galaxies observed through
$r_{ap} < 1$ kpc apertures. All $W$'s are measured with respect to the
Featureless Continuum, whose strength $x_{FC}$ was estimated from our
synthesis analysis. In the H$\beta$ and [OIII] plots the FC strength is
evaluated at 4861 \AA, whereas for [OII] we evaluate $x_{FC}$ at 3660 \AA.}
\label{fig:W_hist}
\end{figure}

This expectation is totally confirmed in the left panels of
Fig.~\ref{fig:W_hist}, where we compare the distribution of
$W^{FC}_{H\beta}$ for our Seyfert 2 sample with that of the 35 Starburst
galaxies in the Starburst sample. {\it These histograms reveal a clear
offset in $W^{FC}_{H\beta}$ between Seyfert 2's and stellar powered
emission line sources}. The median $W^{FC}_{H\beta}$ is a factor of 5
larger in the Seyfert 2 sample than in pure Starbursts (129 compared to 26
\AA).

Given their dual nature, we intuitively expect composite systems to
exhibit $W^{FC}_{H\beta}$ in between those of ``pure'' Seyfert 2's and
Starbursts.  This expectation is also confirmed. The composite Seyfert
2/starburst systems in our sample, marked by the filled region in
Fig.~\ref{fig:W_hist}a, tend to populate the low end of the
$W^{FC}_{H\beta}$ distribution in Seyfert 2's, overlapping with the
high $W^{FC}_{H\beta}$ Starbursts (Fig.~\ref{fig:W_hist}b).

The same considerations, of course, apply to other emission lines.  In
fact, given that [OIII]/H$\beta$ is much larger in Seyferts than in
stellar powered systems, we expect an even clearer separation between
these two types of objects in terms of $W^{FC}_{[OIII]}$.  This is
confirmed in Figs.~\ref{fig:W_hist}c and d. The difference between
median $W^{FC}$'s, which was a factor of 5 for H$\beta$, is now
32-fold, with 44 \AA\ for Starburst galaxies and 1431 \AA\ in the
Seyfert 2 sample. As for H$\beta$, composites are skewed towards low
$W^{FC}_{[OIII]}$'s (Fig.~\ref{fig:W_hist}c). The same effects are
identified in [OII] (Figs.~\ref{fig:W_hist}e and f).

The large offset in $W^{FC}$'s between Seyfert 2's and Starburst galaxies
is not an artifact of aperture differences between the two samples. One
does not expect drastic changes in the line per continuum photon ratio
between nuclear and off-nuclear star-forming regions, so the distribution
of $W^{FC}$'s for Starburst galaxies should remain roughly the same for
small and large apertures.  This is demonstrated by the shaded portions of
the histograms in Fig.~\ref{fig:W_hist}b, d and f, which mark Starburst
galaxies observed through physical apertures $r_{ap} < 1$ kpc.  Aperture
effects become important for Seyfert 2's, since there the contribution of
off-nuclear HII regions affects the line/FC proportion, diluting the higher
$W^{FC}$ of the nucleus. This effect is detected for the objects in common
between the Seyfert 2 and NLAGN samples. Mrk 477, for instance, has
$W^{FC}_{H\beta} = W^{obs}_{H\beta} / x_{FC} = 35 / 0.47 = 75$ \AA\ for the
observations of SBMQ95, whereas with our nuclear spectra we obtain $92 /
0.87 = 107$ \AA. The same effect occurs comparing the nuclear with the
``whole aperture'' spectra of LK95, which integrate over the entire galaxy.

\begin{figure}
\plotone{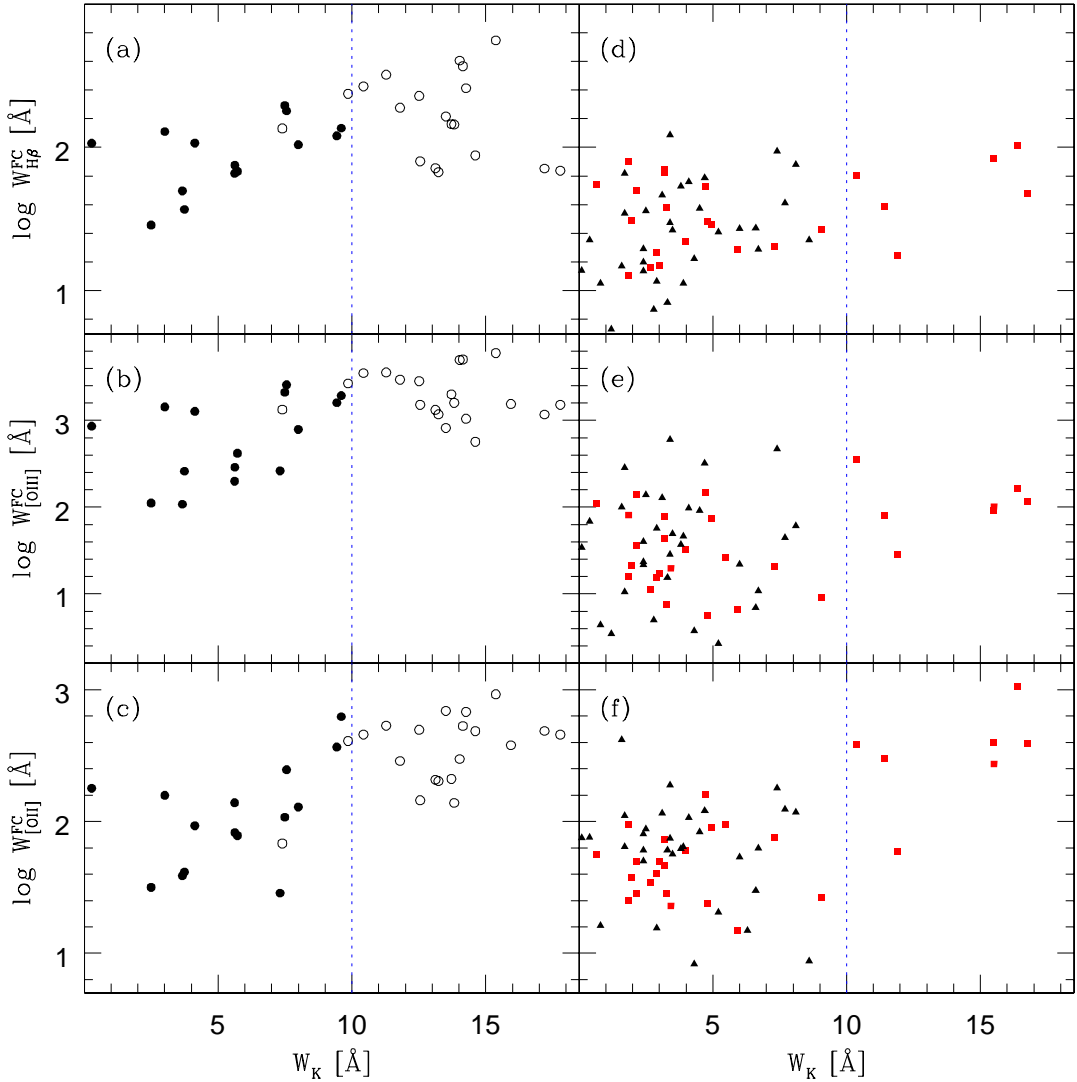}
\caption{Equivalent widths of H$\beta$ (top), [OIII] (middle) and
[OII] (bottom) with respect to the FC against $W_K$. Sources from the
Seyfert 2 sample are in the left (a--c), whereas plots in the right
(d--f) show the Starburst and Merger samples.  Symbols follow the same
convention as in Fig.~\ref{fig:Fig_xFC_X_xINT}.  The location of
Seyfert 2's in the $W_K \times W^{FC}_{H\beta}$ diagram (a) can be
used to assess the {\it degree of difficulty to identify starburst
features}: Sources to the lower left (weak $W_K$ and
$W^{FC}_{H\beta}$) have a strong, starburst dominated FC, whereas we
do not know with certainty what dominates the weak FC sources in the
top right of the diagram.}
\label{fig:Ws_X_WK}
\end{figure}

Since in \S\ref{sec:Synthesis} we found that composites tend to have a
strong FC, we may expect them to be well separated from ``pure''
Seyfert 2's in diagrams involving $x_{FC}$ and these emission line
$W^{FC}$'s. This is illustrated in Figs.~\ref{fig:Ws_X_WK}a--c, where
we plot the $W^{FC}$'s of H$\beta$, [OIII] and [OII] against $W_K$,
here representing the FC strength. Note that all composites are
located to the left of the $W_K = 10$ \AA\ line indicated in the
plots. (A slightly more stringent constraint of $W_K < 8.5$ \AA\ can
be adopted by restraining our definition of composites to systems
which exhibit signs of young stars, thus excluding the
post-starbursts NGC 5643, Mrk 78 and ESO 362-G8.)  The $W^{FC}$'s by
themselves do not segregate composite from ``pure'' Seyfert 2's as
efficiently as $W_K$, but sources with the weakest emission lines are
predominantly composites, so the vertical axis provides some extra
diagnostic power.  The ``pure'' Seyfert 2 at $W_K = 7.4$ \AA,
trespassing the $W_K < 10$ \AA\ zone of composites, is NGC 1068, which
has $W^{FC}_{[OII]} = 68$, $W^{FC}_{[OIII]} = 1334$ and
$W^{FC}_{H\beta} = 136$ \AA.  This prototype Seyfert 2, which shows no
signs of recent star-formation close to the nucleus, is located either
among or close to the composites in all diagrams presented in this
paper. We will return to this issue in \S\ref{sec:Discussion}.
The panels on the right side of Fig.~\ref{fig:Ws_X_WK} show the results for
sources in the Starburst and Merger samples for comparison.

\subsubsection{The difficulty in identifying circum-nuclear starbursts}

\label{sec:Difficulty_to_detect_starbursts}

Once again we must draw attention to contrast effects which may lead
us to classify Seyfert 2's with weak circum-nuclear starbursts as
``pure'' systems. Part of the difficulty in identifying starburst
features in Seyfert 2 nuclei with strong emission-lines comes from the
fact that as $W^{FC}_{H\beta}$ increases the high order HI Balmer
absorption lines of massive stars become increasingly filled up by
emission.  In the absence of far-UV spectra (available for very few
sources) and the short-lived WR bump, this effect effectively prevents
us from seeing the starburst in the near-UV to optical range.

This contrast effect comes on top of the difficulty in identifying
circum-nuclear starbursts in systems dominated by the old stellar
population (\S\ref{sec:Constrast_and_Evolution}). The combination of
these two effects may be at least partly responsible for the
horizontal and vertical separations between composites and ``pure''
Seyfert 2's seen in Fig.~\ref{fig:Ws_X_WK}a--c.  The location of a
Seyfert 2 in the $W_K$-$W^{FC}_{H\beta}$ plane can therefore be read
as an empirical measure of the {\it ``degree of difficulty''} in the
recognition of circum-nuclear starbursts in Seyfert 2's: Sources at
the bottom left are easily identified as composites, whereas as one
progresses to the top right it becomes increasingly difficult to
discern the starburst features. The usefulness of this concept is
illustrated by the location of Mrk 3, Mrk 34 and NGC 7212 in
Figs.~\ref{fig:Ws_X_WK}a--c, all are around the $W_K = 10$ \AA\
``dividing line'' but with $W^{FC}_{H\beta}$ and $W^{FC}_{[OIII]}$
larger than for composites.  It is no accident, therefore, that the
nature of their moderately strong FC remains unclear
(\S\ref{sec:moderate_and_weak_FC_sources}).

\subsection{The relation between line and FC emission}

\label{sec:line_FC_relation}

\begin{figure}
\plotone{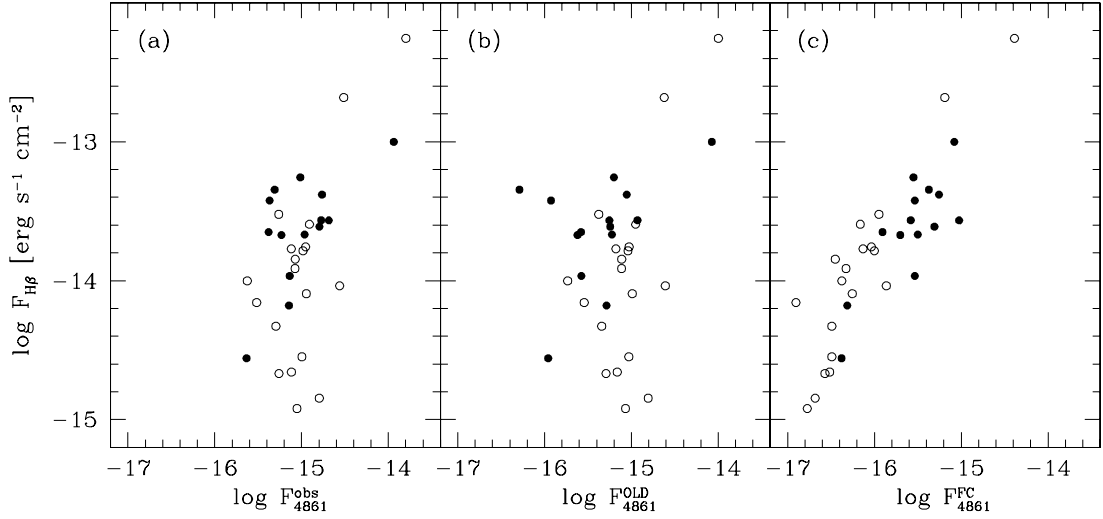}
\caption{Relation between the continuum at 4861 \AA\ and H$\beta$
fluxes.  The panels differ in which component of $F_{4861}$ is used in
the abscissa. (a) Total observed continuum. (b) $F_{4861} \times
x_{OLD}$, i.e., the continuum flux due to old stars. (c) $F_{4861}
\times x_{FC}$, the FC flux.  Symbols as in
Fig.~\ref{fig:Fig_xFC_X_xINT}. All continuum fluxes are given in
erg$\,$s$^{-1}\,$cm$^{-2}\,$\AA$^{-1}$.  Notice how $x_{FC}$ has the
property of unveiling a strong line-continuum correlation immersed in
the scatter plot on the left.}
\label{fig:Hb_x_Continuum}
\end{figure}

It is important to emphasize that the offset in emission line $W$'s
between Seyfert 2's and Starbursts only appears when these are
measured with respect to the FC, as hinted in
Fig.~\ref{fig:Ws_X_WK}. Our estimate of $x_{FC}$, which takes into
consideration continuum colors and metal absorption lines through our
semi-empirical synthesis analysis, is thus the key behind this
significant off-set.

The role of the synthesis process can be better appreciated in
Fig.~\ref{fig:Hb_x_Continuum}. In Fig.~\ref{fig:Hb_x_Continuum}a we
plot $F_{H\beta}$ against the total observed continuum flux at 4861
\AA. What is seen is a scatter diagram, which is not surprising since
the continuum in most sources is dominated by old stars and thus has little
relation to the line emission. This is confirmed by the fact that the
scatter increases even further if the continuum fluxes are multiplied by
$x_{OLD}$ (Fig.~\ref{fig:Hb_x_Continuum}b), thus isolating the flux due to
$\ge 1$ Gyr stars.  The line versus continuum plot changes dramatically
when we isolate the FC contribution!  Using $F_{FC} = F^{obs}_{4861} \times
x_{FC}$ in the abscissa (Fig.~\ref{fig:Hb_x_Continuum}c) has the remarkable
effect of uncovering an underlying order immersed in the scatter plot on
the left. It is therefore clear that {\it our FC determination method
isolated a component of the optical continuum which is directly linked to
the line emission}, lending further credibility to the inferred $x_{FC}$
values. The same result is obtained for the sources in our comparison
samples, i.e., $x_{FC}$ organizes the data along a $F_{H\beta}$-$F_{FC}$
correlation which is not evident in the $F_{H\beta}$-$F^{obs}_{4861}$ plane
because of the contribution of the old stellar population.

\begin{figure}
\plotone{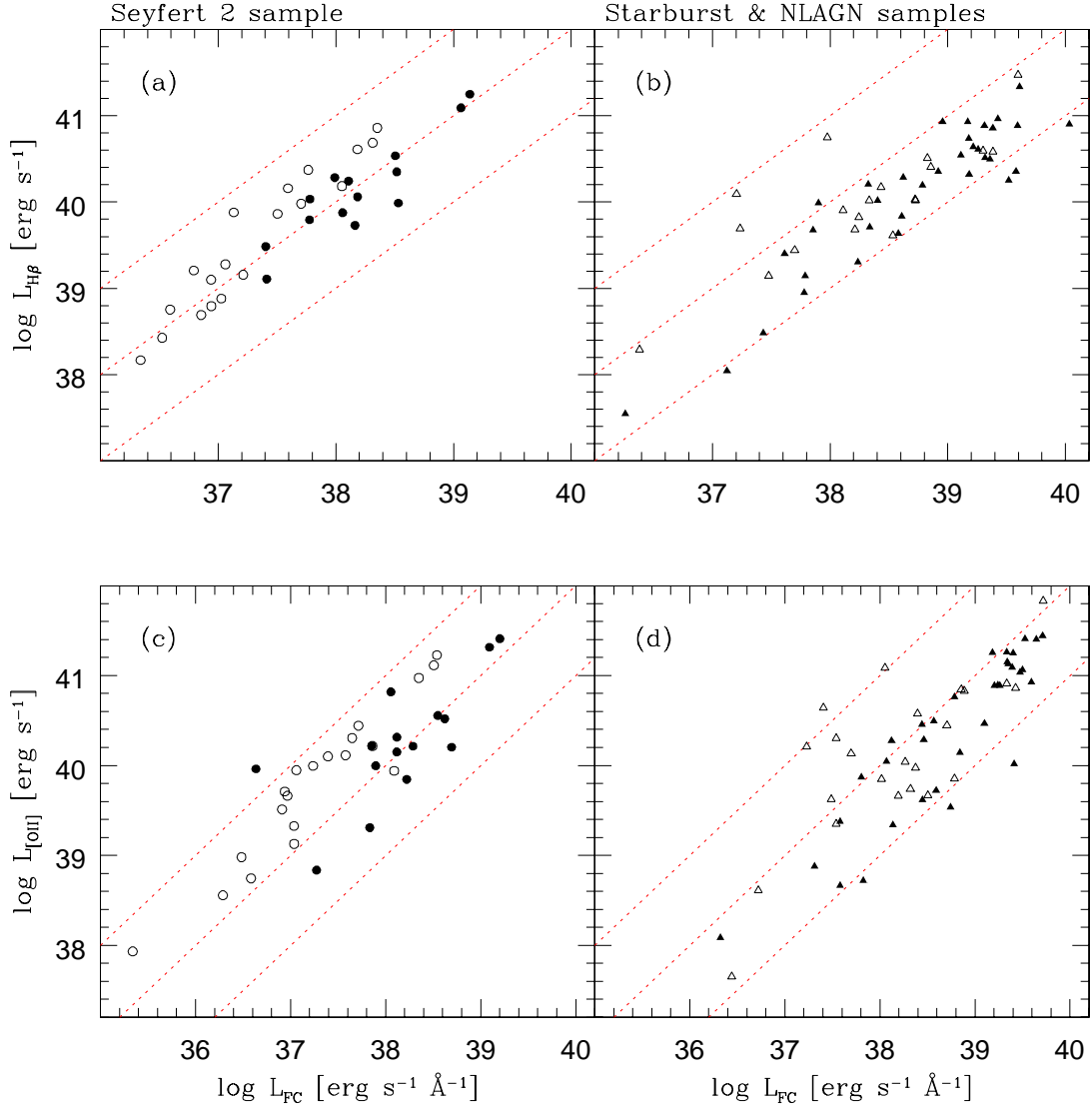}
\caption{Relation between the H$\beta$ (top) and [OII] (bottom)
luminosities and the FC luminosity $L_{FC} = L_{obs} \times x_{FC}$
for the Seyfert 2 (left), Starburst and NLAGN samples (right). Note
that $L_{FC}$ is evaluated at 4861 \AA\ for the H$\beta$ plot and at
3660 \AA\ for [OII].  Diagonal lines indicate, from bottom to top,
$W^{FC} = 10$, 100 and 1000 \AA. Symbols as in
Fig.~\ref{fig:Fig_xFC_X_xINT}.}
\label{fig:Hb_and_O2_X_FC}
\end{figure}

\begin{figure}
\plotone{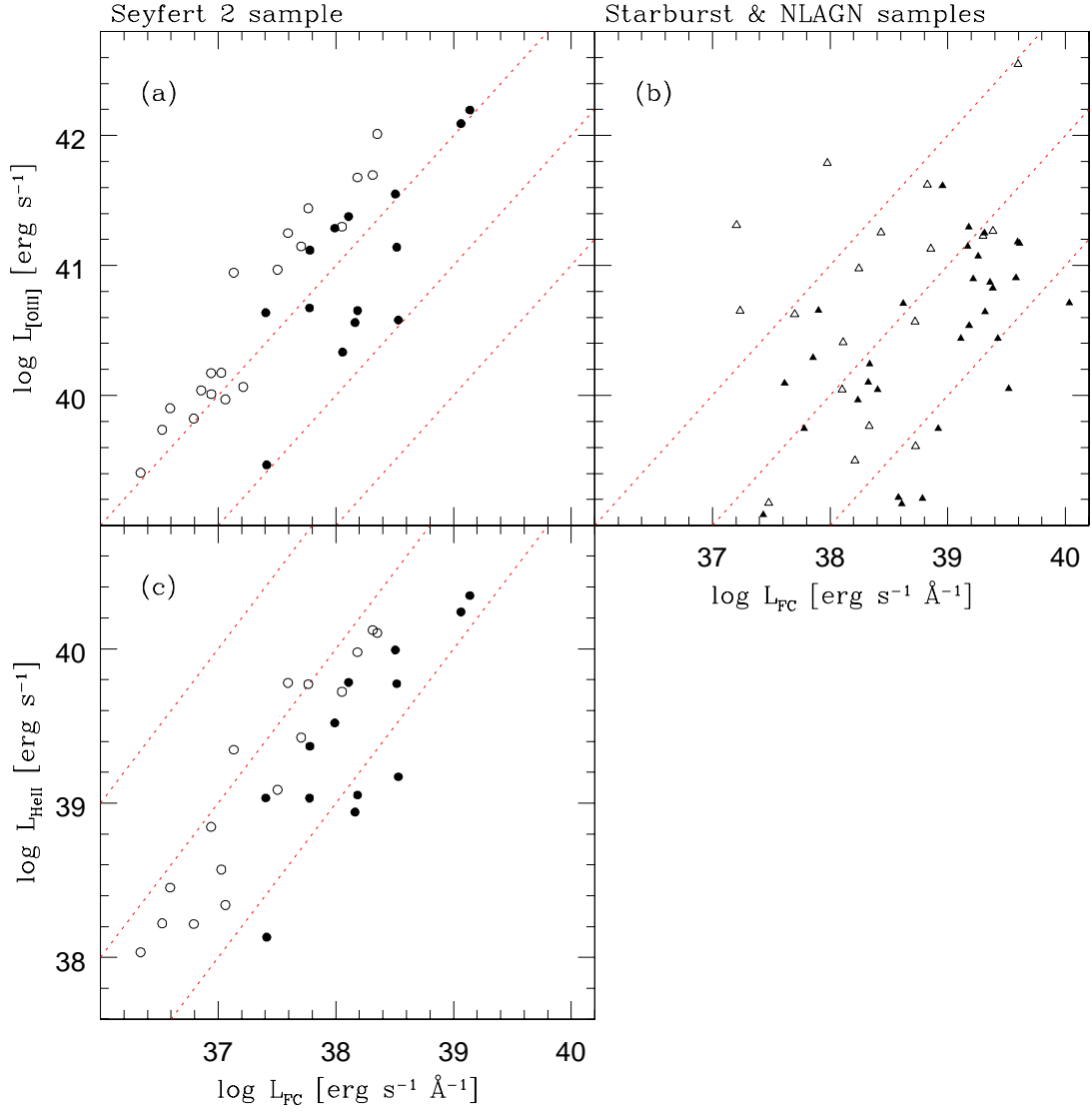}
\caption{Same as Fig.~\ref{fig:Hb_and_O2_X_FC}, but for the [OIII] and
HeII emission lines. No HeII measurements are available for the
Starburst and NLAGN samples.}
\label{fig:He2_and_O3_X_FC}
\end{figure}

\begin{deluxetable}{lrrrrrrrrrrrr}
\rotate
\tabletypesize{\scriptsize}
\tablecaption{Line-FC correlations}
\tablewidth{0pc}
\tablehead{
 & \multicolumn{4}{c}{Whole Sample} &
   \multicolumn{4}{c}{Composites}   &
   \multicolumn{4}{c}{``Pure'' Seyfert 2's} \cr
\colhead{}&
\colhead{$F_{obs}$}&
\colhead{$F_{OLD}$}&
\colhead{$F_{SB}$}&
\colhead{$F_{FC}$}&
\colhead{$F_{obs}$}&
\colhead{$F_{OLD}$}&
\colhead{$F_{SB}$}&
\colhead{$F_{FC}$}&
\colhead{$F_{obs}$}&
\colhead{$F_{OLD}$}&
\colhead{$F_{SB}$}&
\colhead{$F_{FC}$}}
\startdata
$F_{H\beta}$  & $7\ET{-2}$ & $7\ET{-1}$ & $2\ET{-7}$ & $6\ET{-11}$ &
                $5\ET{-1}$ & $6\ET{-1}$ & $3\ET{-1}$ & $2\ET{-2}$  &
                $2\ET{-1}$ & $3\ET{-1}$ & $9\ET{-5}$ & $2\ET{-6}$ \cr
$F_{HeII}$    & $3\ET{-2}$ & $9\ET{-2}$ & $5\ET{-3}$ & $2\ET{-3}$  &
                $-9\ET{-1}$ & $7\ET{-1}$ & $-6\ET{-1}$ & $-5\ET{-1}$ &
                $1\ET{-2}$ & $2\ET{-2}$ & $1\ET{-3}$ & $7\ET{-4}$ \cr
$F_{[OIII]}$  & $2\ET{-1}$ & $5\ET{-1}$ & $1\ET{-3}$ & $4\ET{-5}$  &
                $-8\ET{-1}$ & $9\ET{-1}$ & $-6\ET{-1}$ & $9\ET{-1}$ &
                $2\ET{-1}$ & $3\ET{-1}$ & $5\ET{-4}$ & $3\ET{-5}$ \cr
$F_{[OII]}$   & $1\ET{-3}$ & $1\ET{-1}$ & $1\ET{-4}$ & $2\ET{-6}$  &
                $3\ET{-1}$ & $8\ET{-1}$ & $5\ET{-1}$ & $3\ET{-2}$ &
                $3\ET{-3}$ & $2\ET{-2}$ & $3\ET{-5}$ & $3\ET{-5}$ \cr
$F_{FIR}$     & $4\ET{-2}$ & $7\ET{-1}$ & $3\ET{-5}$ & $1\ET{-4}$  &
                $9\ET{-2}$ & $3\ET{-1}$ & $4\ET{-2}$ & $5\ET{-1}$ &
                $2\ET{-1}$ & $3\ET{-1}$ & $5\ET{-2}$ & $5\ET{-3}$ \cr
\enddata
\tablecomments{Probability $p_S$ of chance flux-flux correlation in a
Spearmann rank test (the smaller $p_S$ the better the
correlation). Negative probabilities denote anti-correlations.  $F_{obs}$
denotes the total optical continuum at 4861 \AA\ (3660 \AA\ for the correlation
with [OII]). $F_{OLD}$, $F_{SB}$ and $F_{FC}$ are the fluxes associated
to the $x_{OLD}$, $x_{INT} + x_{FC}$ and $x_{FC}$ fractions, respectively.}
\label{tab:Correlations}
\end{deluxetable}

Possible interpretations of the H$\beta$-FC relation identified in
Fig.~\ref{fig:Hb_x_Continuum}c will be discussed in
\S\ref{sec:Line_X_FC_Interpretation}, after we gather other related
results. To substantiate the discussion it is useful to compare it to the
equivalent relation in Starburst systems and to investigate line-FC
relations for other transitions.  This is done in
Figs.~\ref{fig:Hb_and_O2_X_FC} and \ref{fig:He2_and_O3_X_FC}, which contain
luminosity-luminosity versions of Fig.~\ref{fig:Hb_x_Continuum}c for the
Seyfert 2, Starburst and NLAGN samples and for H$\beta$, [OII], [OIII] and
HeII.  Lines of constant $W^{FC} = 10$, 100 and 1000 \AA\ run diagonally
across these plots to facilitate their intercomparison.  Also to aid the
discussion, Table~\ref{tab:Correlations} summarizes the results of a
Spearman rank correlation analysis for flux-flux line-continuum
relations. The values listed are the probabilities of chance correlation
($p_S$), so small values indicate significant correlations.

Significant line-FC correlations are also obtained for [OII], [OIII] and
HeII in the Seyfert 2 sample (Figs.~\ref{fig:Hb_and_O2_X_FC}c,
\ref{fig:He2_and_O3_X_FC}a and \ref{fig:He2_and_O3_X_FC}c,
Table~\ref{tab:Correlations}). Note, however, that the scatter in the
[OIII] and HeII plots is visibly larger than for H$\beta$-FC
(Fig.~\ref{fig:Hb_and_O2_X_FC}a).  It is also clear that a substantial part
of the scatter in these plots is induced by the offset location of the
composites towards smaller $W^{FC}$'s. Statistical confirmation of this
effect is provided in Table~\ref{tab:Correlations}, where these
correlations are examined for the whole sample and for subsets containing
only ``pure'' or composite systems.

Starburst galaxies also follow a tight H$\beta$-FC relation, with
$W^{FC}_{H\beta} = 10$--100 \AA\ (filled triangles in
Figs.~\ref{fig:Hb_and_O2_X_FC}b). This is a long known (Terlevich \etal
1991) and well understood behavior of stellar powered systems, for which
$F_{FC}$ actually represents the contribution of young stars and hence
traces the ionizing flux. Current models (e.g., Leitherer \etal 1999)
predict $W_{H\beta}$ between 400 and 10 \AA\ for constant star-formation,
different IMF's and ages up to $10^8$ yr. These are broadly consistent with
the values we obtain, but a detailed comparison with such evolutionary
models will have to await a cross calibration of our semi-empirical
synthesis decomposition in terms of the theoretical spectra.  As for the
Seyfert 2 sample, the second best line-FC relation for Starburst galaxies
is that with the [OII] line (Fig.~\ref{fig:Hb_and_O2_X_FC}d). This too is a
well known relation, which has in fact been used as an alternative
estimator of star-formation rates in galaxies (Gallagher \etal 1989;
Kennicutt 1992; Tresse \etal 1999; Jansen, Franx \& Fabricant 2001).

The location of the NLAGN's in these plots (empty triangles) reflects
the mixed nature of this sample. Some of them behave just like
``pure'' Seyfert 2's, most notably NGC 3393 and NGC 5506, which stand
out in Figs.~\ref{fig:Hb_and_O2_X_FC}b, d and
\ref{fig:He2_and_O3_X_FC}b as the sources with highest
$W^{FC}$'s. Most of the other NLAGN's are mixed among Starburst galaxies in
these plots. This happens due to a combination of the aperture effects
discussed in \S\ref{sec:Ws} and the fact that this sample contains several
composites and some objects which are more consistent with a LINER
classification (e.g., NGC 1097, NGC 1433). It is thus clear that these
galaxies do not constitute a well defined comparison sample.  For these
reasons, we concentrate on the comparison between the Seyfert 2 and
Starburst sample.

\subsubsection{Extinction}

\label{sec:extinction}

\begin{figure}
\plotone{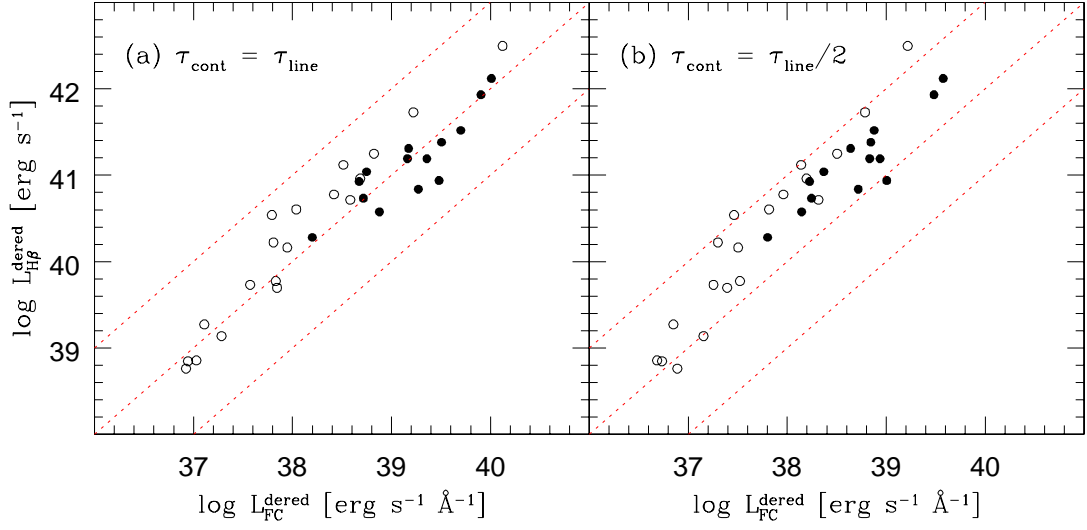}
\caption{(a) Dereddened FC versus H$\beta$ relation
for the Seyfert 2 sample after correcting both axis for the extinction
implied by H$\alpha$/H$\beta$. (b) As above, but assuming the continuum is
only half as extincted as the emission lines (Calzetti \etal 1994).
Symbols and lines as in Fig.~\ref{fig:Hb_and_O2_X_FC}.}
\label{fig:Hb_X_FC_reddening}
\end{figure}

One would like to believe that extinction does not play a relevant role in
these line-FC diagrams since both axes correspond to similar wavelengths.
In Starburst galaxies, however, it is known that the continuum suffers
$\sim 2$ times less extinction than the emission line region (Calzetti,
Kinney \& Storchi-Bergmann 1994). This effect can reshuffle the data points
in the plots above, potentially changing the overall appearance and
strength of the line-FC correlations.

An estimate of the effects of extinction is presented in
Fig.~\ref{fig:Hb_X_FC_reddening} for the H$\beta$-FC relation using
the H$\alpha$/H$\beta$ values listed in Table~\ref{tab:TABELAO2}.
Fig.~\ref{fig:Hb_X_FC_reddening}a presents the data with $L_{FC}$ and
$L_{H\beta}$ dereddened by the same amount, whereas in
\ref{fig:Hb_X_FC_reddening}b we halved the optical depth to the
FC. Two effects are clear when comparing these plots with their
uncorrected counterpart (Fig.~\ref{fig:Hb_and_O2_X_FC}a): First, with
the extinction correction, composites are more segregated from
``pure'' Seyfert 2's in terms of $L_{H\beta}$, with median values
which differ by an order of magnitude (median $L_{H\beta} =
1.6\ET{41}$ and $1.7\ET{40}$ erg$\,$s$^{-1}$ for composites and
``pure'' Seyfert 2's respectively). This preference of composites for
luminous systems will appear again when we discuss their far IR
luminosities in \S\ref{sec:IRAS}. Second, a differential line/FC
extinction correction only tightens the H$\beta$-FC correlation! The
effect is such that it practically erases the offset in
$W^{FC}_{H\beta}$ between composite and ``pure'' Seyfert 2's.

The caveats in these results are that the H$\alpha$/H$\beta$ values
used were not obtained from the same spectra we analyze
(\S\ref{sec:Sample}) and that we do not know whether the $\tau_{FC}
\sim \tau_{H\beta} / 2$ result applies to AGN. At any rate, the larger
H$\alpha$/H$\beta$ of composites, which is responsible for this regrouping,
is consistent with UV (Heckman \etal 1995) and X-ray (Levenson, Heckman \&
Weaver 2001a) indications that circum-nuclear starbursts in Seyfert 2's are
substantially reddened, and may significantly contribute to the extinction
of the AGN component in such composite systems.  This ``excess'' of dust is
probably associated with the same gas which feeds the starburst activity,
but more work will be required to address this issue quantitatively.

\subsection{Do starbursts participate in the ionization of the gas in
Seyfert 2's?}

\label{sec:He2_and_O3}

\begin{figure}
\plotone{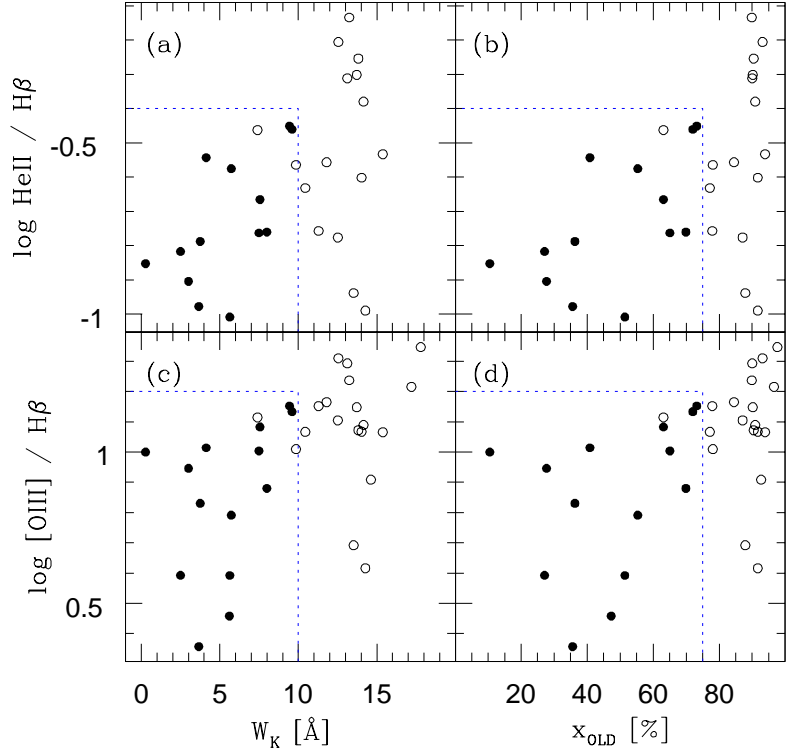}
\caption{``Excitation'' ratios plotted against indicators of
circum-nuclear starbursts for the Seyfert 2 sample.  Filled circles
mark composite systems and empty circles mark ``pure'' Seyfert 2's.
Dotted lines delimit regions occupied by composites, and may be used
as empirical diagnostics of compositeness.}
\label{fig:He2Hb}
\end{figure}

The results reported so far show how the presence of circum-nuclear
starbursts manifest itself in the FC strength and the equivalent
widths of emission lines. These results {\it per se} do not
necessarily imply that the starburst contributes to the ionization of
the line emitting gas in Seyfert 2's, since it may act mostly as a
source of FC, diluting an AGN powered H$\beta$ emission and thus
shifting points to the right, but not much upwards in
Figs.~\ref{fig:Hb_x_Continuum}---\ref{fig:Hb_X_FC_reddening}.  One way
to establish whether the starburst participates in the ionization of
the gas is to investigate the ratio between HeII$\lambda$4686 and
H$\beta$. Unlike AGN, starbursts do not produce significant
radiation above 54 eV, so they show little HeII emission. If
starbursts in AGN contribute to H$\beta$ this ratio will tend to be
smaller in composites than in ``pure'' Seyfert 2's.

This is confirmed in Figs.~\ref{fig:He2Hb}a--b, where we plot
HeII/H$\beta$ against $W_K$ and $x_{OLD}$, both empirical indicators
of compositeness (\S\ref{sec:Synthesis}).  Whereas the whole sample
spans the 0.1--0.8 interval in HeII/H$\beta$, all composites have
HeII/H$\beta < 0.4$.  Many ``pure'' Seyfert 2's also have HeII this
weak, but their stronger Ca II K and hence larger $x_{OLD}$ clearly
separates them from composites.  Starburst galaxies would be located
in the same horizontal range spanned by composites in
Fig.~\ref{fig:He2Hb}a--b, but down to weaker HeII (a line which is
often not detected in starbursts). Composites also tend to be less
excited in terms of [OIII]/H$\beta$, as shown in
Figs.~\ref{fig:He2Hb}c--d.  The effect of the stellar absorption
feature underneath the H$\beta$ emission-line was not corrected for in
this analysis. Since composites are more likely to have stronger
Balmer absorption lines than objects dominated by old stars, this
correction would only strengthen our conclusion, shifting the
composites to even smaller HeII/H$\beta$ and [OIII]/H$\beta$ ratios.

Either line ratio can therefore be used as an auxiliary diagnostic of
compositeness.  This is illustrated by the dotted lines in
Fig.~\ref{fig:He2Hb}, which isolate composites from ``pure'' Seyfert
2's almost completely! (Again, the odd ball is NGC 1068.)  This shows
a welcome consistency between diagnostics of compositeness based on
continuum and stellar features and the expected impact of
circum-nuclear starbursts upon the emission line ratios in Seyfert
2's.  Note, however, that while all known composites have low
excitation, the converse is not true, since many ``pure'' Seyfert 2's
lie in the HeII/H$\beta < 0.4$ and [OIII]/H$\beta < 15$ range defined
by composites.  In terms of emission line ratios, these galaxies could
well have circum-nuclear starbursts, but which do not make it to the
composite class due to the contrast effect discussed above.

It is hard to see how to account for these results in a ``pure AGN''
scenario. The shape of the ionizing spectrum, the proportion of matter
to ionization bounded clouds and the ionization parameter are just
some of the factors which define the excitation level of the NLR, and
source-to-source variations of these properties can account for a wide
spread in HeII/H$\beta$ and [OIII]/H$\beta$ (e.g. Viegas-Aldrovandi
1988). Yet, none of these intrinsic properties explains the ``zone of
avoidance'' for strong FC and high excitation in
Fig.~\ref{fig:He2Hb}. If anything, a strong FC would suggest a higher
ionization parameter and perhaps a harder ionizing spectrum, which
would produce higher excitation, contrary to what is
observed. Overall, the simplest interpretation of these data is that
part of the emission lines in composites originates from
photoionization by a circum-nuclear starburst. This contribution is
strongest in H$\beta$, since starbursts are inefficient producers of
HeII and [OIII] compared to AGN.

It is clear that in a starburst $+$ AGN mixture of emission lines the
starburst component can dominate H$\beta$ without moving the source
outside of the AGN region in diagnostic diagrams (e.g., Hill \etal
2001; Levenson \etal 2001b).  For instance, using typical
[OIII]/H$\beta$ ratios of 10--20 for Seyfert 2's and 0.3--1 for
metal-rich starbursts, the total ratio still lies within the
[OIII]/H$\beta > 3$ domain of Seyferts for as much as 70--90\% of
H$\beta$ powered by the starburst!  Therefore our conclusion that
photoionization by massive stars provides a substantial part of
$L_{H\beta}$ in composites is not in conflict with their Seyfert 2
classification. In fact, this conclusion was already implicit in the
off-set in $W^{FC}_{H\beta}$ between composites and Starburst galaxies
(Fig.~\ref{fig:W_hist}).  The $W^{FC}_{H\beta}$ values in composites
cover the range from 30 to 200 \AA, which may be represented by its
geometric mean of $\sim$80 \AA. The corresponding value in Starburst
galaxies is $\sim$30 \AA, representing the $W^{FC}_{H\beta} = 10$--100
\AA\ interval.  Since we know that the FC in composites is dominated
by the starburst component, these values imply a typical starburst
contribution of $30/80 \sim 40\%$ to H$\beta$.

These results imply that H$\beta$\ is partially powered by the
starburst in composite systems. Unfortunately, as for $L_{FC}$, it is
not possible to unambiguously separate the starburst and AGN shares of
$L_{H\beta}$ without extra information, such as assuming typical line
ratios or equivalent widths. For instance, taking the mean
[OIII]/H$\beta$ of 13 for our ``pure'' Seyfert 2's and 8 for
composites, one obtains a starburst contribution to H$\beta$ of 40\%
(precisely what we have derived above comparing composites and
Starburst galaxies in terms of $W^{FC}_{H\beta}$).  Illustrative as
these global estimates are, it is clear that both composite and
``pure'' systems present a {\it range} of properties.  For the
composites below NGC 5135 ([OIII]/H$\beta = 4$) in
Fig.~\ref{fig:He2Hb}c and d, for instance, the same exercise would
yield more than 70\% of H$\beta$ powered by the starburst.  Much
therefore remains to be learned from a more detailed object by object
analysis.

\subsection{Emission Line Profiles: Kinematical separation of the
starburst and AGN}

\label{sec:LineProfiles}

One prospect for disentangling the starburst and AGN contributions to
the emission lines is to study their profiles in search of
differential kinematical signatures of these two components. We have
thus performed a line profile analysis for the Seyfert 2 sample, but
only for the Kitt Peak data, which have higher spectral resolution.

Most of the composites have [OIII] and H$\beta$ profiles which can
each be described in terms of two components. Interestingly, the
narrower component often has a lower excitation than the broader one
(Table \ref{tab:Profiles}), qualitatively consistent with them being
powered by a starburst and AGN respectively. In fact, the excitation
of the broad component (mean [OIII]/H$\beta = 10.7$, standard
deviation $= 2.8$) is very similar to the excitation of ``pure''
Seyfert 2's. However, the excitation of the narrow components ranges
between 2.7 and 11.6. Meanwhile, the excitation in Mrk 1066, Mrk 1073,
NGC 5135, NGC 7130 and IC 3639 is similar to the excitation in
prototypical nuclear starbursts (e.g.\ NGC 7714, Gonz\'alez-Delgado
\etal 1995), while in Mrk 1, Mrk 463E, Mrk 477 and Mrk 533 the
excitation is higher and similar to the excitation in young and very
highly excited HII regions (e.g.\ NGC 2363, Gonz\'alez-Delgado \etal
1994). The difference in excitation between these two sub-groups
within the composites is suggestive of evolutionary effects, where the
youngest systems are the ones with highest excitation. This is
consistent with other properties of composites, such as
$W^{FC}_{H\beta}$, the fraction of total optical light provided by
intermediate and young stars ($x_{INT}/x_{FC}$), and the detection of
Wolf-Rayet features (see \S\ref{sec:Evolution}).

If this interpretation is correct, i.e., the narrow and broad emission
line components are produced by the starburst and AGN respectively,
the ratio of the narrow to total line flux gives an estimate of the
starburst contribution to the ionization of the gas. For H$\beta$,
this ratio ranges between 37\%\ (in Mrk 463E) and 80\%\ (in Mrk 1073
and NGC 5135), in agreement with our coarser estimates in
\S\ref{sec:He2_and_O3}.  These values suggest a significant impact of
the circum-nuclear starbursts upon the emission line ratios in Seyfert
2s.

\begin{deluxetable}{lrrrr}
\footnotesize
\tablecaption{Line Profile Analysis of Composites}
\tablewidth{0pt}
\tablehead{
\colhead{Name} &
\colhead{$\left(\frac{\rm [OIII]}{H\beta}\right)_{narrow}$} &
\colhead{$\left(\frac{\rm [OIII]}{H\beta}\right)_{broad}$} &
\colhead{$\frac{F_{H\beta,narrow}}{F_{H\beta}}$ [\%]} &
\colhead{$\frac{F_{[OIII],narrow}}{F_{[OIII]}}$ [\%]} \cr
}
\startdata
Mrk1       &   11.6     &      11.5    & 38 & 38 \cr
Mrk273 & 3.1\tablenotemark{\star} & & & \cr
Mrk463E    &   8.9      &      8.3     & 37 & 39 \cr
Mrk477     &   8.8      &      8.7     & 58 & 59 \cr
Mrk533     &   9.4      &      12.5    & 54 & 46 \cr
Mrk1066    &   2.9      &      5.7     & 70 & 55 \cr
Mrk1073    &   4.6      &      15.4    & 80 & 54 \cr
NGC5135    &   2.7      &      10.5    & 80 & 54 \cr
NGC7130    &   2.7      &      11.7    & 60 & 25 \cr
IC3639     &   5.4      &      12.0    & 66 & 47 \cr
\enddata
\tablenotetext{\star}{The nuclear spectrum of Mrk 273 corresponds to
the central 2.1$\times$1.5 arcsec around the continuum maximum at 4450
\AA.  This maximum is shifted with respect to the maximum of the
nebular emission lines.  The [OIII] map obtained by Colina, Arribas
\& Borne (1999) shows an off-nucleus Seyfert 2 nebula 4 arcsec south with a
[OIII]/H$\beta \sim 8$, similar to the excitation of the broad component in
the other composites.}
\tablecomments{The principal properties of the two kinematic components
fitted to the H$\beta$ and [OIII] $\lambda$5007 emission lines in the
composites Seyfert 2 nuclei.}
\label{tab:Profiles}
\end{deluxetable}

\subsection{What drives the line-FC correlations in Seyfert 2's?}

\label{sec:Line_X_FC_Interpretation}

The FC strength, $W_{FC}$'s, emission line ratios and profiles all
consistently reflect the presence of circum-nuclear starbursts in
Seyfert 2's and can therefore be used as the empirical diagnostics of
compositeness which we set out to identify in this paper.  The line-FC
correlations presented in Figs.~\ref{fig:Hb_and_O2_X_FC} and
\ref{fig:He2_and_O3_X_FC} do not play any direct part in such
diagnostics. Nevertheless, because of their high statistical
significance, potential relevance and controversial history, in this
section we speculate on what might be driving them.  We do this by
contrasting two extreme views on the origin of the FC in ``pure''
Seyfert 2's.

\subsubsection{Starburst-dominated FC}

Since we have established that circum-nuclear starbursts make substantial
contributions both to the FC and H$\beta$ emission, and given that these
two quantities are causally linked in starburst systems, {\it the presence
of circum-nuclear starbursts in Seyfert 2's naturally leads to a
H$\beta$-FC correlation}. Even if the AGN components of $F_{FC}$ and
$F_{H\beta}$ are uncorrelated, the starburst portions should suffice to
maintain a H$\beta$-FC correlation in the combined AGN $+$ starburst data.

One extreme interpretation of Fig.~\ref{fig:Hb_and_O2_X_FC}a is thus
that the H$\beta$-FC correlation is essentially driven by the presence
of circum-nuclear starbursts. If this is to apply to {\it all}
sources, the FC must be dominated by the starburst {\it even in
``pure'' Seyfert 2's}. Varying proportions of starburst to AGN
ionizing power can be invoked to account for the few ``pure'' Seyfert
2's whose H$\beta$/FC ratios exceed the maximum $W^{FC}_{H\beta} =
200$ \AA\ reached by composites. This can be explained either by
scaling up the nuclear ionizing source, thus increasing $F_{H\beta}$
while keeping $F_{FC}$ constant, or by scaling down the starburst,
which also increases $W^{FC}_{H\beta}$ since $F_{FC}$ would be more
affected than $F_{H\beta}$. Such adjustments can even be dispensed
with if one allows for differential extinction to the line and FC,
since we have seen in Fig.~\ref{fig:Hb_X_FC_reddening}b that this
alone substantially reduces the $W^{FC}_{H\beta}$ differences between
composite and ``pure'' systems.

But how do the [OIII] (Fig.~\ref{fig:He2_and_O3_X_FC}a) and HeII-FC
(Fig.~\ref{fig:He2_and_O3_X_FC}c) correlations fit into this model?
Table~\ref{tab:Correlations} shows that these correlations, while poorer
than the one between $F_{H\beta}$ and $F_{FC}$, are still significant at
the $>99\%$ confidence level, and that the reason for the larger scatter is
the small $W^{FC}_{[OIII]}$ and $W^{FC}_{HeII}$ of some 5 or 6 composites.
Such a scatter is in fact expected in a starburst dominated FC model, since
the [OIII] and (especially) the HeII line fluxes trace the AGN power,
whereas $F_{FC}$ is a measure of the starburst power, so that
$W^{FC}_{[OIII]}$ and $W^{FC}_{HeII}$ are indicators of the varying
contrast between these components. Decreasing the power of the starbursts
in composite systems by factors of 2--3 while keeping the AGN constant
would move composites along $\sim$ horizontal lines to the left in
Figs.~\ref{fig:He2_and_O3_X_FC}a and c, placing them among ``pure'' Seyfert
2's.

While it is easy to see why composites deviate in $W^{FC}_{[OIII]}$ and
$W^{FC}_{HeII}$, it is not clear why ``pure'' Seyfert 2's should define
line-FC correlations as good as those revealed in
Figs.~\ref{fig:Hb_and_O2_X_FC} and
\ref{fig:He2_and_O3_X_FC} (see also Table~\ref{tab:Correlations}). If
their FC is indeed dominated by a weak starburst, then the fact that their
AGN-dominated $F_{[OIII]}$ and $F_{HeII}$ scale with $F_{FC}$ would
indirectly imply the existence of a {\it proportionality between the
starburst and AGN powers}, i.e., that more powerful starbursts occur in
more powerful AGN!  Further evidence for this scaling is presented in
\S\ref{sec:IRAS}, \S\ref{sec:luminosity_connection} and GD01.

\subsubsection{AGN-dominated FC}

An opposite model would be to postulate that the FC is dominated by
starbursts {\it only} in composites, while in ``pure'' Seyfert 2's the FC
is dominated by scattered light, so these two classes should not be mixed
when discussing line-FC relations. This interpretation would lead us to the
intriguing conclusion that the scattering efficiency, $\epsilon$, which
links the observed FC1 to the intrinsic nuclear FC via $L_{FC1} = \epsilon
L_{FC0}$, does not vary substantially among Seyfert 2's. The surprise comes
from the fact that $\epsilon$ depends on the geometry of the mirror and the
optical depth of scattering particles (Miller, Goodrich \& Mathews 1991;
Cid Fernandes \& Terlevich 1995). Since there is no {\it a priori} reason
why $\epsilon$ should be similar for all Seyfert 2's, we would expect a
{\it wide spread} of $W^{FC}$'s, contrary to the relatively narrow ranges
observed, particularly if composites are excluded.

This argument has in fact been pursued by Mulchaey \etal (1994) to explain
the {\it absence} of line-FC correlations in their data, from which they
claim a consistency with the unified model. As acknowledged by Mulchaey
et\thinspace al., however, it is possible that, just as we find here, a
substantial part of the scatter in their line-FC plots is associated with
contamination by starburst activity.  That this is more than a possibility
is illustrated by the fact that the Seyfert 2 list of Mulchaey \etal (1994)
includes 12 of our 15 composites!  It also includes other prime composite
suspects, most notably NGC 6221, a source for which Levenson \etal (2001b)
find a complete dominance of the starburst, with the AGN shining through
only in hard X-rays. As a whole, their sample is similar to the NLAGN
sample of SBMQ95 in terms of the mixed types of sources. Indeed, as seen in
Figs.~\ref{fig:Hb_and_O2_X_FC}b, d and \ref{fig:He2_and_O3_X_FC}b, no
line-FC correlations are present for this heterogeneous sample.
Furthermore, contamination by starbursts is certainly larger in the
Mulchaey \etal data than in ours, since they trace the FC by the far-UV
flux collected through the large aperture of the IUE spectrometer, a
procedure which follows the premise that starlight contamination is
irrelevant in the UV. This premise was proven wrong by Heckman \etal
(1995), who have demonstrated that, on the contrary, scattered light from a
hidden Seyfert 1 is a minor contributor to the IUE spectra of Seyfert 2's,
which are more typical of reddened starbursts.

At least in relative terms, it is clear that our observations are
better suited to investigate line-FC relations. Our results
essentially recover the classic Yee (1980) and Shuder (1981)
correlation between line and FC luminosity, which extends over more
than 5 decades in luminosity from Seyfert 2's to QSOs. Looking at
their original plots one sees that Seyfert 2's show a larger scatter
around the correlations than type 1 sources, but they still broadly
follow the correlations.  While it is true that the interpretation of
their line-FC correlations was left orphan with the advent of the
unified model, their reality has not been convincingly challenged as
yet. Similarly, barring unidentified selection effects, it is hard to
deny the reality of the line-FC correlations seen in
Figs.~\ref{fig:Hb_and_O2_X_FC} and \ref{fig:He2_and_O3_X_FC}.

In a $L_{FC} = L_{FC1}$ model $W^{FC}_{H\beta}$ equals $W^{Sey 1}_{H\beta}
\epsilon^{-1}$, where $W^{Sey 1}_{H\beta}$ is the equivalent width which
would be measured from non-obscured lines of sight to the Seyfert 1
nucleus. Observationally, the narrow H$\beta$ in Seyfert 1's roughly spans
the range between $W^{Sey 1}_{H\beta} \sim 5$ to 75 \AA\ (Goodrich 1989;
Rodr\'{\i}guez-Ardila, Pastoriza \& Donzelli 2000). This spread is already
comparable to the $70 \le W^{FC}_{H\beta} \le 560$ \AA\ interval span by
``pure'' Seyfert 2's in our sample, which does not leave much room for a
spread in $\epsilon$.  Combining these intervals we obtain an acceptable
range for $\epsilon$ of roughly 0.01--0.1. (As a reference, for NGC 1068,
the only source for which the scattering has been modeled in detail,
$\epsilon \sim 0.015$; Miller \etal 1991). A ``well behaved'' $\epsilon$ is
thus required in order to explain the H$\beta$-FC correlation for ``pure''
Seyfert 2's. This interpretation is analogous to, and just as puzzling as,
the interpretation of the constancy of the $W$ of broad H$\beta$ among type
1 Seyferts and QSOs as due to a roughly constant covering fraction of broad
line clouds in all sources (Yee 1980; Shuder 1981; Osterbrock 1989;
Binette \etal 1993). While viable, this model obviously
requires some fine tuning.

Yet another possibility is that the FC carries a sizable {\it nebular}
component in ``pure'' Seyfert 2's, which would naturally produce
line-FC correlations. A purely nebular FC, however, would produce
$W^{FC}_{H\beta}$ of order 2000 \AA\ for electron temperatures between
10000 and 20000 K, much larger than the values we find.  A hotter
plasma, however, can enhance the continuum emission and lower the
$W^{FC}$'s. This is the essence of Tran's (1995c) model for FC2:
emission from the $\sim 10^5$ K electron-scattering mirror. If this
emission is somehow tied to the line emission---which is reasonable
since both arise in circum-nuclear gas heated by the active
nucleus---then a line-FC correlation follows.

While our personal prejudices tend to favor a starburst-dominated FC
interpretation, we are not in a position to rule out a role for a
non-stellar FC for ``pure'' Seyfert 2's. Indeed, the fact that NGC 1068
mingles so well among composite starburst/Seyfert 2's is a clear
demonstration that our analysis cannot resolve the ambiguous nature of the
FC in ``pure'' Seyfert 2's. Since we know that several composites have a
scattered component (Table~\ref{tab:TABELAO2}), the correct interpretation
of the line-FC relation will certainly involve a mixture of the extreme
possibilities discussed above.

\section{Far IR luminosities}

\label{sec:IRAS}

\begin{figure}
\plotone{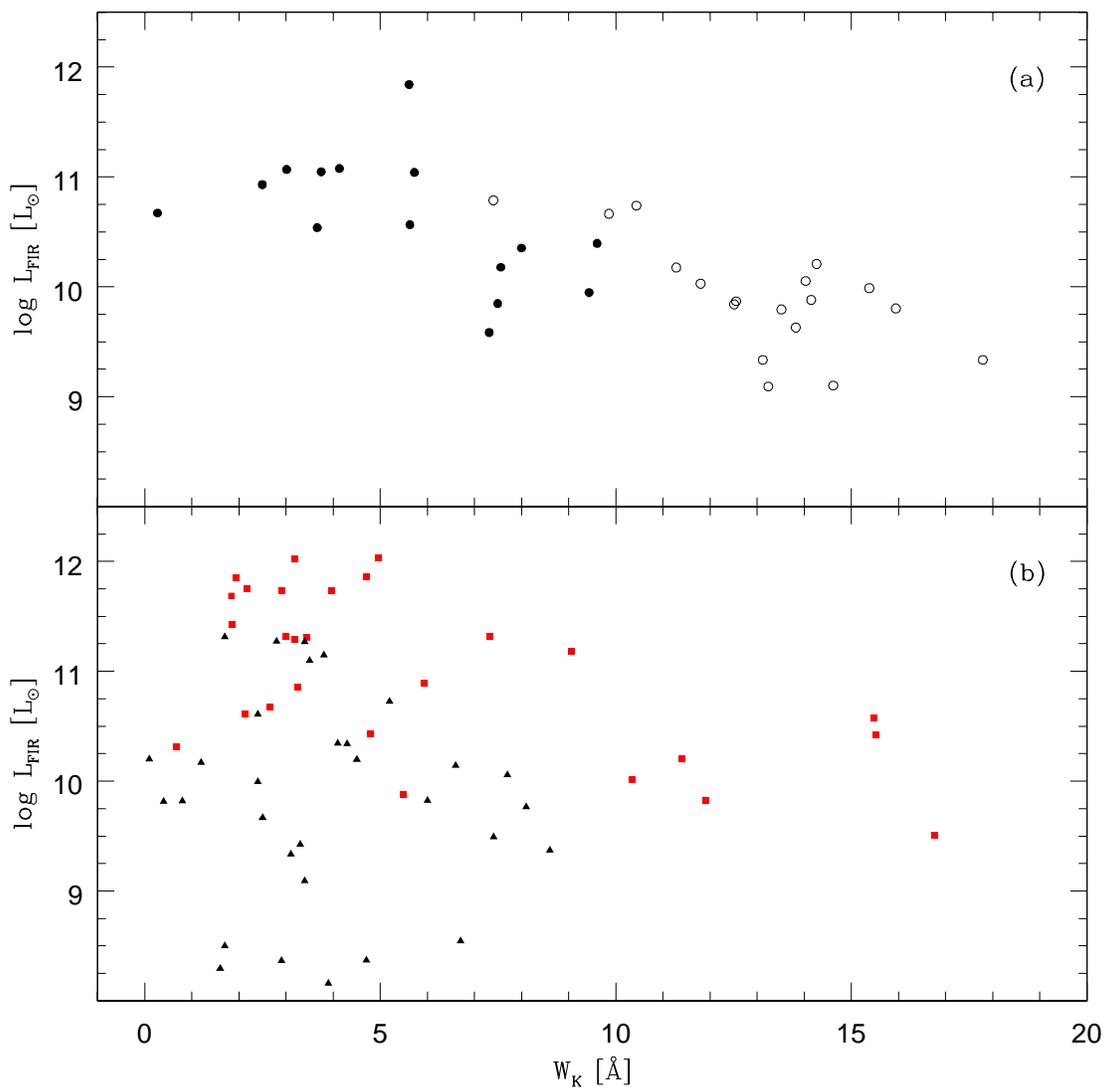}
\caption{Far IR luminosity against $W_K$ for (a) the Seyfert 2 sample,
and (b) the Starburst and Merger samples. Symbols as in
Fig.~\ref{fig:Ws_X_WK}. Starbursts with $L_{FIR} \leq$ few $\times 10^9$
L$_\odot$ are generally blue compact dwarfs or related objects. Normal
galaxies would populate the bottom part of the figure (i.e., low
$L_{FIR}$), with $W_K$ depending on the amount of star-formation. Seyfert
2's lie along the upper envelope, with composites skewed to high
$L_{FIR}$.}
\label{fig:IRAS}
\end{figure}

In Figs.~\ref{fig:Hb_and_O2_X_FC}--\ref{fig:Hb_X_FC_reddening} we have seen
that, besides having smaller emission line $W^{FC}$'s, the tendency of
composites to live preferentially among luminous sources allows them to be
more clearly separated from ``pure'' Seyfert2's in line-FC diagrams. To
evaluate this tendency in a reddening independent fashion,
Fig.~\ref{fig:IRAS}a shows the far IR luminosity, computed with the 60 and
100 $\mu$m IRAS bands (cf.\ Sanders \& Mirabel 1996), against $W_K$ for the
Seyfert 2 sample.  The tendency identified by GD01 for composites to be
powerful IR emitters is clearly seen in this figure. Except for Mrk 1210,
ESO 362-G8 and NGC 5643, all composites have $L_{FIR} > 10^{10}$
L$_\odot$. The difference in median $L_{FIR}$ between composites and
``pure'' Seyfert 2's is a factor of 5.

Comparison with Fig.~\ref{fig:IRAS}b shows that composites are also more
powerful far IR sources than typical Starburst galaxies, being more
comparable to the merger systems studied by LK95.  Normal galaxies with
little star-formation would populated the large $W_K$ and low $L_{FIR}$
portion of the plot (Telesco 1988; Soifer, Houck \& Neugbauer 1987).

IR luminous galaxies have a well know preference to be interacting
systems (Sanders \& Mirabel 1996). Fig.~\ref{fig:IRAS}a therefore
suggests that interactions and the associated gas fueling of the
nuclear regions may be a key ingredient in triggering circum-nuclear
starbursts in AGN.  Indeed, a substantial fraction of our composites
(9 of 15) are associated with interacting systems or groups (GD01;
Levenson \etal 2001a; SB00): NGC 7130, Mrk 1, Mrk 273, Mrk 463E, Mrk
477, Mrk 533 and IC 3639 are all in interacting systems, while NGC
5135 and NGC 7582 belong to groups. The analogous fraction for the the
``pure'' Seyfert 2's is only 4 out of 20 (Mrk 348, Mrk 607, NGC 5929
and NGC 7212). This may explain why the composites resemble the
mergers in Fig.~\ref{fig:IRAS}.

\begin{figure}
\plotone{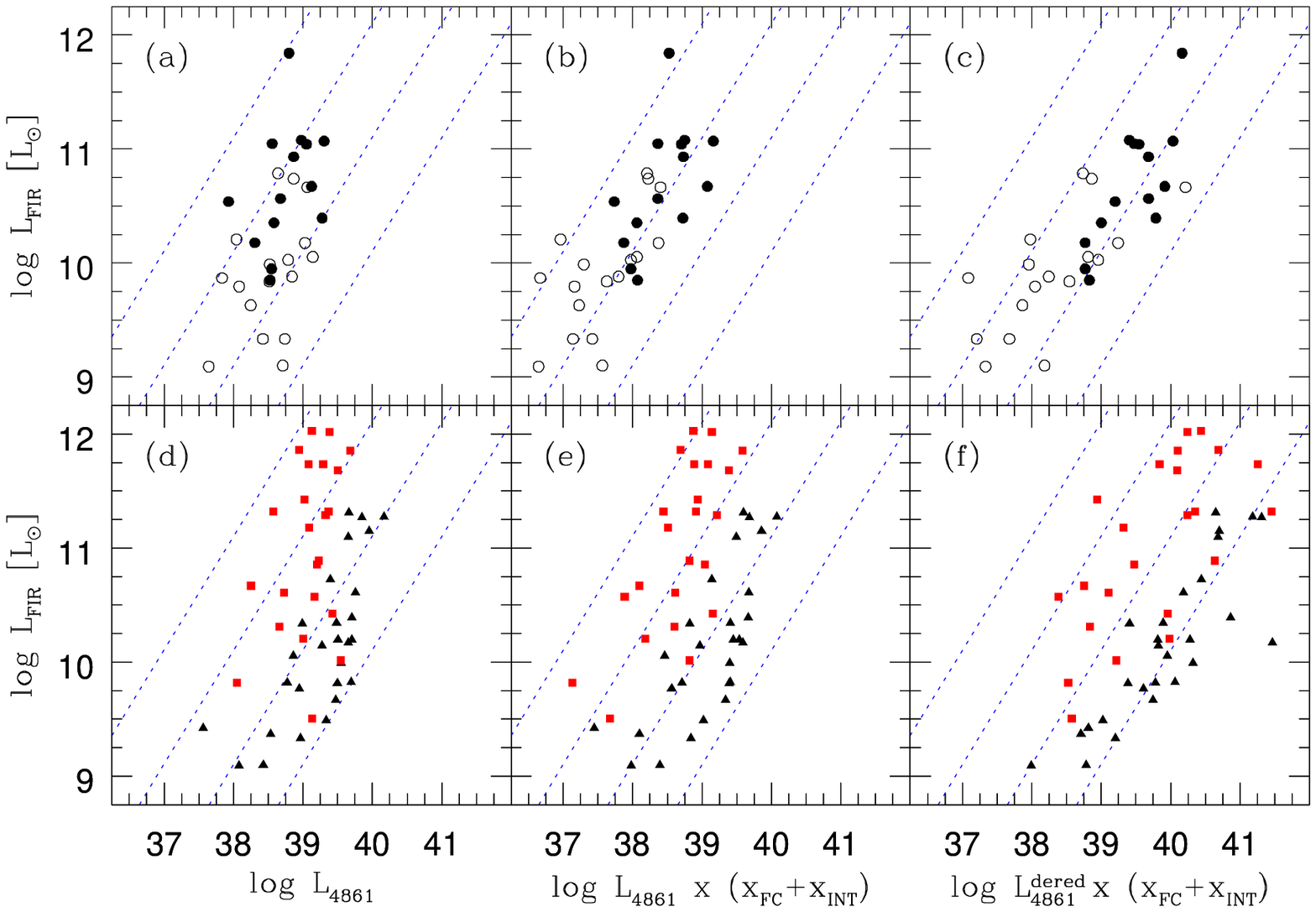}
\caption{{\it Top:} Relation between the far-IR and the continuum luminosity
at 4861 \AA\ for the Seyfert 2 sample. In (a) the total $L_{4861}$ is used
in the abscissa, whereas only the FC $+$ post-starburst component is used
in (b), and in (c) $L_{4861} \times (x_{FC} + x_{INT})$ is corrected for
extinction (using H$\alpha$/H$\beta$). Composites are marked by filled
circles.  {\it Bottom:} As above, but for the Starburst (filled triangles)
and Merger (squares) samples. Abscissa in units of
erg$\,$s$^{-1}\,$\AA$^{-1}$. The dotted lines correspond to $L_{FIR} / (\nu
L_\nu)_{4861}$ ratios of 1000, 100, 10 and 1.}
\label{fig:IRAS_x_FC}
\end{figure}

Since $W_K$ works essentially as a measure of the starburst strength,
Fig.~\ref{fig:IRAS} indicates a link between the far IR and the amount of
recent star-formation in Seyfert 2's.  This link is investigated in
Fig.~\ref{fig:IRAS_x_FC}.  The abscissa in Fig.~\ref{fig:IRAS_x_FC}a is the
total observed luminosity at 4861 \AA, whereas in Fig.~\ref{fig:IRAS_x_FC}b
we filter out the light from old stars by using $L_{SB} \equiv
L^{obs}_{4861} \times (x_{FC} + x_{INT})$. The post-starburst population,
which was not included in the line-FC analysis because it does not ionize
the gas, is now included because it can heat the dust.  Furthermore,
Table~\ref{tab:Correlations} shows that the far IR flux is somewhat better
correlated with $F_{SB}$ than with $F_{FC}$ alone. Comparison of
Figs.~\ref{fig:IRAS_x_FC}a and b show how $x_{FC}+x_{INT}$ has the property
of unveiling a correlation immersed in the scatter when the whole continuum
is used. An estimate of the effects of extinction is presented in
Fig.~\ref{fig:IRAS_x_FC}c, where we deredden $L_{SB}$ using the
H$\alpha$/H$\beta$ values listed in Table~\ref{tab:TABELAO2}. The bottom
panels in Fig.~\ref{fig:IRAS_x_FC} show the corresponding results for
Starburst galaxies and mergers. Comparison of the top and bottom panels
shows that Seyfert 2's have $F_{FIR}/F_{SB}$ ratios larger than galaxies in
the Starburst sample, but compatible with the LK95 mergers.  Despite the
mixed emission line properties of the sources in the Merger sample, it is
reasonable to attribute most of their FIR and FC emission to star-formation
in a dusty environment (Lutz \etal 1999; 1998). This similarity is thus
suggestive of the presence of powerful dusty circum-nuclear starbursts in
Seyfert 2's, consistent with our knowledge of composites. A usual caveat
when dealing with IRAS fluxes is that one has to allow for the fact that a
substantial fraction of the far IR light originates well outside the
nuclear region. Aperture effects may therefore be partially responsible for
the larger $F_{FIR}/F_{SB}$ found for objects in the Seyfert 2 than in the
Starburst sample. This effect was detected for sources in common between
the Seyfert 2 and NLAGN samples, the latter having systematically larger
$F_{SB}$ and thus smaller $F_{FIR}/F_{SB}$.  At any rate, it is clear that
circum-nuclear starbursts can be responsible for the bulk of the far IR
emission in Seyfert 2's.  This was in fact confirmed in the few cases where
this issue has been examined more closely (e.g., Mrk 477 Heckman \etal
1997; see also Rigopoulou \etal 1999).

\section{Near UV surface brightness}

\label{sec:SurfBrightness}

\begin{figure}
\plotone{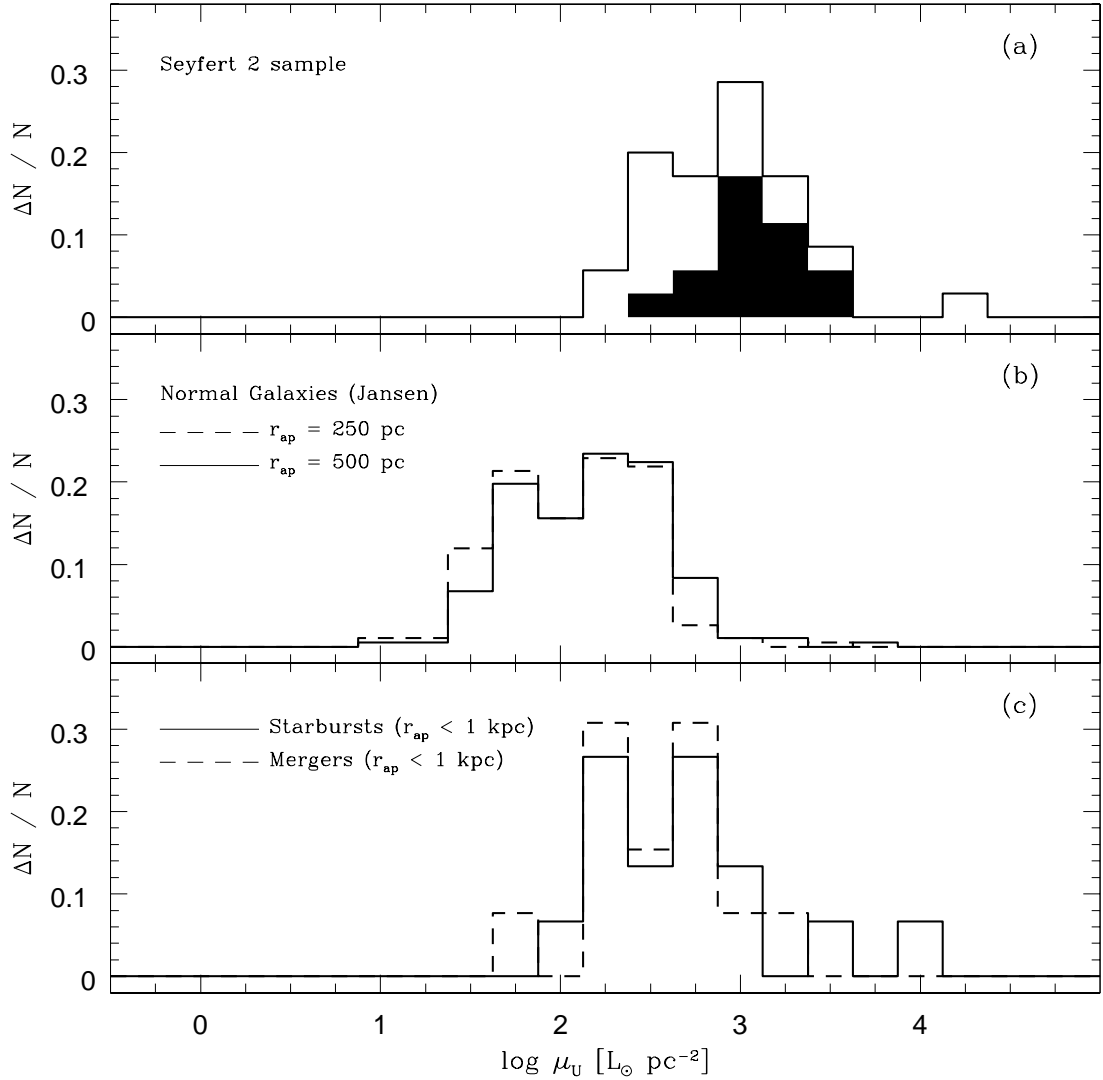}
\caption{(a) Distribution of U-band surface brightness for the Seyfert 2
sample, with composites indicated by the filled areas.  (b) As above, but
for normal galaxies from the Jansen \etal (2000) catalog, extracted through
250 or 500 pc aperture radii.  (c) As above but for the Starburst and
Merger samples, excluding sources observed through apertures larger than
$r_{ap} = 1$ kpc.  For the Seyfert 2, Starburst and Merger samples $\mu_U$
was measured from the $F_{3660}$ flux divided by the aperture area.  For
the Jansen \etal data $\mu_U$ comes from aperture photometry in U.}
\label{fig:SBU1}
\end{figure}

\begin{figure}
\plotone{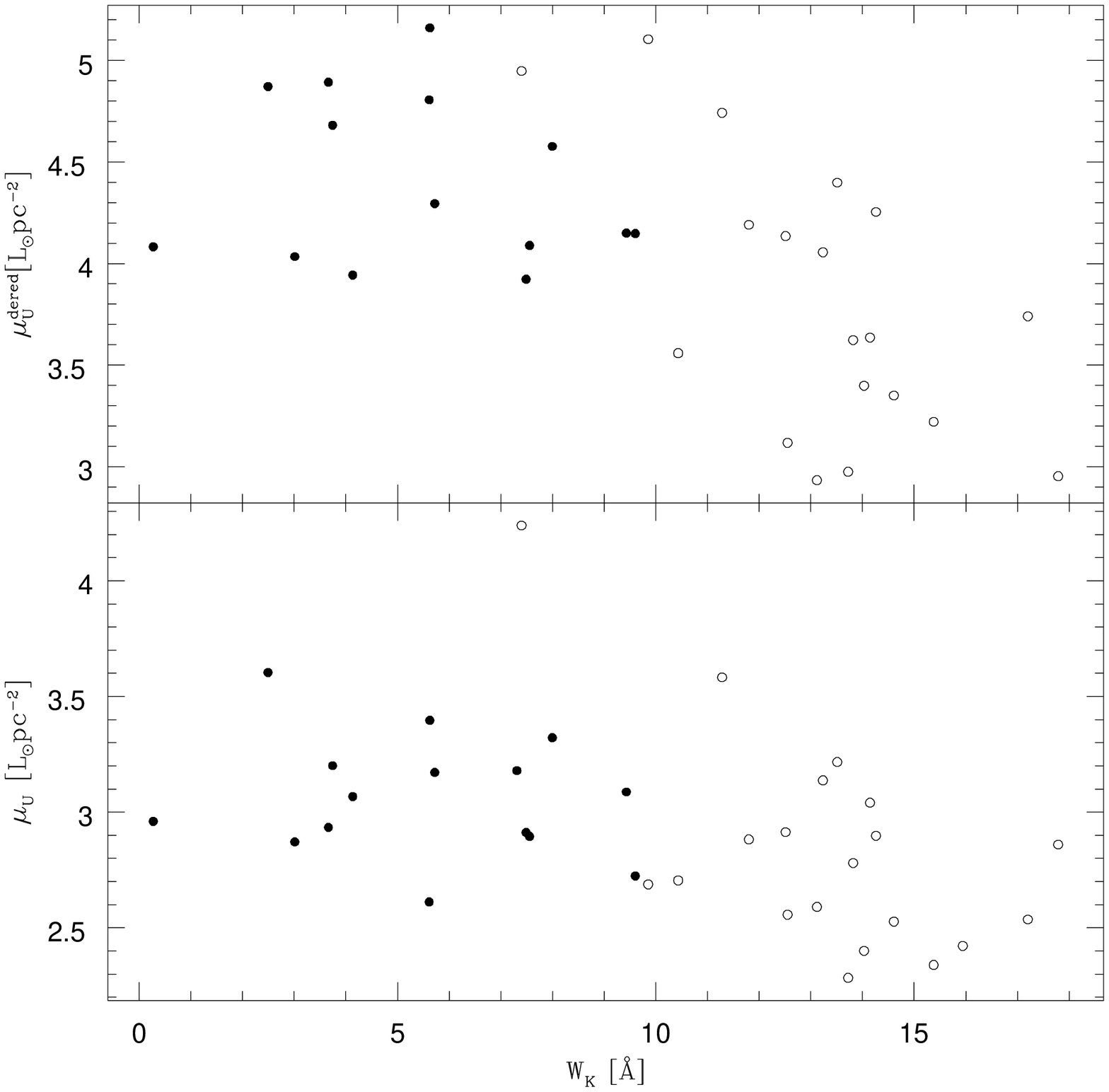}
\caption{{\it Bottom:} U-band surface brightness against $W_K$ for the
Seyfert 2 sample. {\it Top:} Same as above after correcting $\mu_U$ by
the extinction using H$\alpha$/H$\beta$. Considering that the
extinction in the continuum can be smaller than that of the emission
lines (Calzetti \etal 1994), panels a and b bracket the correct values
for $\mu_U$.}
\label{fig:SBU2}
\end{figure}

We have shown that the near-UV nuclear light in the Seyfert 2
composites is dominated by young and intermediate-age stars.  In order
to make some causal connection between these young stars and the
Seyfert phenomenon, we must establish that the properties of this
stellar population are unusual compared to the nuclei of normal
galaxies. Young and intermediate-age stars dominate the near-UV
continuum in the nuclei of normal galaxies of Hubble type Sbc and
later (e.g. Heckman 1980).  Seyfert galaxies are generally classified
as early type disk galaxies (e.g. Nelson \& Whittle 1996; Ho,
Filippenko \& Sargent 1997), whose nuclei are dominated by old stars.
However, the Hubble types of Seyfert galaxies based on ground-based
images are quite uncertain, since they are generally much more distant
than the galaxies used to define the Hubble sequence. Indeed, the HST
imaging survey by Malkan, Gorjian \& Tam (1998) reveals a significant
number of Seyfert 2's (including some of our composites) in which the
HST-based Hubble type is much later than that assigned from earlier
ground-based images (Storchi-Bergmann \etal 2001). Could it be that
the Seyfert 2 composites are simply otherwise normal late-type
galaxies containing an AGN? The abnormally large far-IR luminosities
of the composites argues against this. In this section we also show
that the observed UV surface-brightnesses of the Seyfert 2 composite
nuclei are much larger than the nuclei of normal galaxies (of any
Hubble type), but rather similar to typical starbursts.

We have compared the surface brightnesses in the near UV ($\sim 3660$
\AA) of all our Seyfert 2's with those of normal galactic nuclei based
on the imaging data published by Jansen \etal (2000). For the normal nuclei
(spanning the full range of Hubble types) we have computed the mean surface
brightness through two metric apertures with radii of 250 and 500 pc,
chosen to represent typical values of the spectroscopic apertures used for
the Seyfert 2 nuclei (see Fig.~\ref{fig:Aperture_X_WK}).  This was done
converting the U-band surface photometry of Jansen \etal (2000) to
monochromatic surface brightnesses at 3650 \AA, including a correction for
Galactic extinction (from NED). We have also measured near-UV surface
brightnesses from the spectra of the Starburst and Merger samples. In these
samples we have restricted our comparison to galaxies in which the
projected aperture size of the spectra had a radius less than 1 kpc (see
Fig.~\ref{fig:Aperture_X_WK}). Thus, in all cases we are comparing UV
surface-brightnesses on similar physical scales (radii of several hundred
pc).

The results are shown in Fig.~\ref{fig:SBU1}, from which we draw the
following conclusions. First, the Seyfert 2 composites have much
higher central UV surface brightnesses than normal galaxies of any
Hubble type (by roughly an order-of-magnitude on average). A
correction for dust extinction would undoubtedly increase the
disparity still further. Second, the Seyfert 2 composites have rather
similar central UV surface-brightnesses to the members of the
Starburst and Merger samples. When the UV surface-brightness is
corrected for extinction (see Fig.~\ref{fig:SBU2}), the Seyfert 2
composites reach values similar to those which would be exhibited by
the ``maximal'' starbursts studied by Meurer \etal (1997).  Third, the
Seyfert 2 composites have tend to have higher central UV
surface-brightness than the ``pure'' Seyfert 2's (by an average factor
of about 2).  As we show in Fig.~\ref{fig:SBU2}, the difference
between the two classes is larger (factor of $\sim$ 4) when an
extinction correction is made.  Fourth, the ``pure'' Seyfert 2's
overlap with the normal galactic nuclei having the highest UV surface
brightnesses. These latter tend to be the nuclei of normal early-type
(Sb and earlier) galaxies whose high UV surface brightness is due to
their prominent bulge component. This is consistent with ``pure''
Seyfert 2's being otherwise normal early-type galaxies with only a
modest contribution of light from an AGN or young stars (as shown by
our spectral synthesis).  Note that the outlying Seyfert 2 with
extremely high surface brightness is the nucleus of NGC 1068 (due to
the small physical size of the spectroscopic aperture and the strong
contribution by scattered AGN light). Finally, Fig.~\ref{fig:SBU2}
also shows the expected trend between increasing UV surface brightness
and an increasing relative contribution to the near-UV light by young
and intermediate-age stars (as parameterized by $W_K$).

\section{Discussion}

\label{sec:Discussion}

\subsection{Diagnostic of compositeness}

Our experiments have consistently shown that Seyfert 2's with
circum-nuclear starbursts detected by means of either far-UV stellar
wind lines, high order Balmer absorption lines or the WR bump also
present distinctive optical/near-UV continuum, optical emission line,
far IR continuum, and near-UV surface brightness properties, all in
agreement with the expected impact of starbursts upon such
observables. Given this remarkable consistency, one can use the
results of our analysis as alternative tools to identify composite
sources.

The following empirical criteria can be used to characterize a Seyfert
2 nucleus as composite:

\begin{enumerate}

\item Weak Ca II K: $W_K < 10$ \AA. Or, equivalently, a small
contribution of old stars: $x_{OLD} < 75 \%$.

\item Weak emission lines with respect to the FC: $W^{FC}_{H\beta} <
200$ \AA, $W^{FC}_{[OIII]} < 2500$ \AA, $W^{FC}_{[OII]} < 600$ \AA.

\item Low excitation: HeII/H$\beta < 0.4$, [OIII]/H$\beta < 15$.

\item Large far-IR luminosity: $L_{FIR} \ga 10^{10}$ L$_\odot$.

\item Composite line profiles, with a narrow component typical of HII-nuclei
and a broader one with NLR-like excitation.

\item High near-UV surface-brightnesses on circum-nuclear scales: $\sim
10^3$ L$_{\odot}$ pc$^{-2}$ uncorrected for extinction and $\sim 10^4$
L$_{\odot}$ pc$^{-2}$ after extinction correction.

\end{enumerate}

These criteria are laid roughly in order of importance, the Ca II K (or
$x_{OLD}$) limit being by far the most useful. The upper limits for the
$W^{FC}$'s and emission line-ratios quoted above were chosen as to
encompass all composites in our sample, most of which have substantially
smaller values than those listed above. As argued above, one cannot base a
classification on these properties alone, but they are useful to confirm
the compositeness for sources satisfying the first criterion.  Taken
together, these empirical rules separate composite from ``pure'' systems
very efficiently.

Naturally, there are border line cases. The moderate FC ($15\% < x_{FC} <
30\%$) galaxies Mrk 3, Mrk 34 and NGC 7212, which we discussed in
\S\ref{sec:moderate_and_weak_FC_sources}, lie either within or
close to the limits outlined above.  The clearest distinction between these
galaxies and our composites is in terms of their emission line $W$'s,
$W^{FC}_{H\beta}= 235$--321 and $W^{FC}_{[OIII]} = 2667$--3565 \AA, typical
of ``pure'' Seyfert 2's but larger than in any of our composites, and well
away from Starburst galaxies.  Note, however, that these high $W^{FC}$'s,
coupled with the dominant contribution of old stars ($x_{OLD} \sim 77\%$)
are precisely the conditions under which the identification of starburst
features are most difficult (\S\ref{sec:Difficulty_to_detect_starbursts}),
so we cannot rule out the possibility that their FC is dominated by
relatively weak circum-nuclear starbursts.  The other moderate FC source in
our sample is NGC 1068, for which there is overwhelming evidence that the
nuclear optical FC is predominantly scattered AGN light. The similarities
between this prototype Seyfert 2 and our composites, whose FC are dominated
by starbursts, is a reminder of how similar starbursts and AGN can be, and
a demonstration of the limitations of our diagnostics for border line
cases.

Overall, the empirical criteria listed above fulfill our objective of
delineating a region in a space of easily obtained observables which
can be used to indirectly infer the presence of circum-nuclear
starbursts in Seyfert 2's. We therefore anticipate that their
application to large samples of near-UV spectra will reveal many more
composite systems.  In fact, since these criteria mix several kinds of
properties, it is not at all obvious whether they will hold for larger
samples.  Future applications of these diagnostics will therefore not
only serve as a means of identifying composite candidates but to
verify the consistency of the scheme as a whole.  It will also be
interesting to incorporate other multi-wavelength data in order to
help disentagling the starburst and AGN components.  Valuable tools in
this sense were developed for the IR part of the spectrum by Miley,
Neugebauer \& Soifer (1985), Lutz \etal (1998), Genzel \& Cesarsky
(2000), Laurent \etal (2000) and Hill \etal (1999, 2001), among
others. In the radio part of the spectrum, some of the diagnostics and
techniques presented by Kewley \etal (2000), Smith \etal (1998a,b),
Lonsdale \etal (1993) and Condon \etal (1991) can be used, while
X-rays signatures of circumnuclear starbursts have been recently
investigated by Levenson \etal (2001a).

\subsection{A ``Starburst $\propto$ AGN'' connection?}

\label{sec:luminosity_connection}

We have shown that starburst/Seyfert 2 composites have large H$\beta$,
FC and far IR luminosities within our sample. These are all quantities
which may carry a significant, even dominant contribution from the
starburst itself.  In order to investigate whether these powerful
starbursts are associated with correspondingly powerful AGN we
compared the [OIII], HeII and 12 $\mu$m luminosities of composites and
pure Seyfert 2's. These properties are less affected by the starburst
component and thus provide an empirical measure of the AGN intensity.

We find that composites are also skewed towards large luminosities in
all these properties.  For instance, 53\% of the sources above but
just 31\% of those below the median reddening-corrected $L_{[OIII]}$
of $4.8\ET{41}$ erg$\,$s$^{-1}$ are composites. Nearly identical
numbers apply to $L_{HeII}$, while for $L_{12\mu m}$, which arguably
measures the hot dust heated by the AGN (Pier \& Krolik 1992;
Spinoglio \& Malkan 1989), the composite percentages are 65 and 25\%
above and below the median respectively.  These asymmetries are
comparable to those in starburst-sensitive properties. For
$L_{H\beta}$, for instance, we find 65\% of composites above the
median $6.0\ET{40}$ erg$\,$s$^{-1}$, but just 19\% bellow it.  The
corresponding numbers for $L_{FIR}$ are 71 and 19\%.

Circum-nuclear starbursts in Seyfert 2's, at least those we can
detect, are therefore powerful (bolometric luminosities of $\ga
10^{10}$ L$_{\odot}$, judging by $L_{FIR}$), and surround
correspondingly powerful AGN. This {\it luminosity link} between AGN
and circum-nuclear starburst activity can be interpreted in several
ways. Perhaps the starburst activity stirs up the ISM in a way which
induces a large rate of gas accretion by the central super-massive
black-hole (Colina \& Wada 2000), while less powerful AGN can be feed
with more modest accretion rates, dispensing the helping hand of a
circum-nuclear starburst. While feasible, this does not seem to be the
only mechanism to produce powerful AGN, since there are also some
high luminosity ``pure'' Seyfert 2's in our sample.  Or perhaps the
``$L_{starburst} \propto L_{AGN}$'' relation extends all over the
$L_{AGN}$ spectrum, but is only detected for the most luminous AGN
because of our difficulties to identify weak circum-nuclear
starbursts.

Whatever the correct interpretation, our results strongly suggest that
powerful Seyfert 2's have at least a 1 in 2 chance of harboring detectable
circum-nuclear starbursts. This {\it prediction} can be readily tested by
applying the diagnostics developed in this and our previous work to near-UV
spectra of such systems, some of which have luminosities rivaling those of
quasars (de Grijp \etal 1992). Extending the starburst-AGN connection
studies towards the low luminosity end, on the other hand, will require
alternative techniques. High spatial resolution (i.e., HST) near-UV
spectroscopy offers the best hope of maximizing the contribution of
circum-nuclear star-forming regions with respect to the old stellar
population background. This technique was recently applied to the LINER NGC
3507 by Shields \etal (2000), where young stars previously undetected from
the ground were clearly revealed with HST.

\subsection{The meaning of compositeness}

\label{sec:Evolution}

\begin{figure}
\plotone{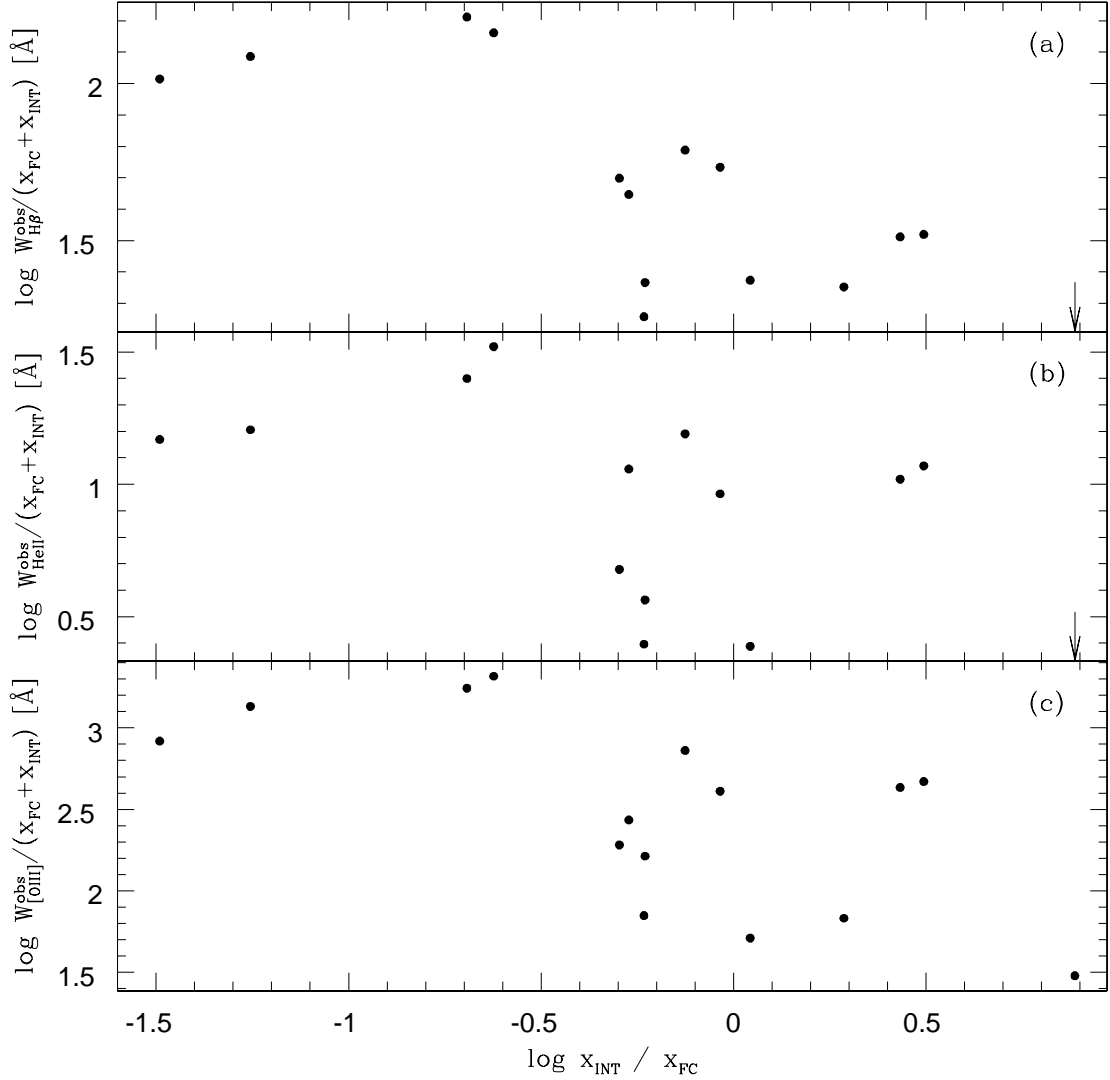}
\caption{(a) ``Evolution'' of the H$\beta$ equivalent width with respect to the
continuum flux due to young and intermediate age stars for the composites
in the Seyfert 2 sample.  The $x_{INT}/x_{FC}$ axis is an empirical age
indicator, which roughly maps onto a 0 to a few $\times 10^8$ yr age
interval within the limits above. (b) and (c) As above, but for the AGN
sensitive HeII and [OIII] lines.  The arrow at the bottom right of (a) and
(b) marks the horizontal position of ESO 362-G8, for which we did not
measure H$\beta$ nor HeII emission.}
\label{fig:Evolution}
\end{figure}

The composite {\it vs.} ``pure'' Seyfert 2 classification which runs
throughout this paper does not necessarily reflect an intrinsic
dichotomy among type 2 Seyferts.  In the spirit of unification, one
would rather prefer to look at these two classes as representing
extremes of a continuous distribution of some underlying physical
parameter.  This view is warranted by the continuity and overlap in
observed properties between composites and ``pure'' Seyfert 2's found
in this and other studies (Levenson \etal 2001a; Veilleux 2001 and
references therein).

At least two ingredients must be involved in any physical description
of a starburst-AGN connection, one of which is some measure of the
relative starburst to AGN intensities, such as the ratio of
star-formation to accretion rate, or the black-hole to stellar mass
ratio.  Of the properties investigated in this paper the $W^{FC}$'s,
mainly of [OIII] and HeII, and the excitation ratios [OIII]/H$\beta$
and HeII/H$\beta$ are in principle the ones more closely associated
with this ``starburst-to-AGN ratio''. The fact that as a whole
``pure'' Seyfert 2's behave more like AGN in terms of these indicators
than composites is thus consistent with them having a smaller
starburst-to-AGN ratio.  The strongest contrast effect, however, is
between the starburst and old stellar population, as deduced from the
fact that {\it all} composites but only one ``pure'' Seyfert 2 have
$x_{OLD} < 75 \%$, a limit which can be expressed in a more empirical
way by $W_K < 10$ \AA. The Ca II K line originates in bulge stars,
being neither a tracer of starburst nor AGN activity, so this
threshold is clearly an observational limitation which prevents us
from identifying circum-nuclear starbursts not powerful enough to
dilute $W_K$ below 10 \AA.  The ``pure'' category ought to contain at
least some such systems, and in fact several ``pure'' Seyfert 2's in
our sample, particularly the less luminous ones (see
Fig.~\ref{fig:Hb_and_O2_X_FC}), have $W^{FC}$'s and excitation ratios
within the range spanned by composites.

The other ingredient is {\it evolution}. The combination of
starburst-to-AGN ratios and evolution can in principle account for the
whole range between composite and ``pure'' systems.  While our coarse,
3-component population synthesis analysis does not provide a detailed
description of the evolutionary history of circum-nuclear starbursts
in Seyfert 2's, in \S\ref{sec:Constrast_and_Evolution} we have seen
that the relative strengths of $x_{FC}$ and $x_{INT}$ in composites
bear a very good correspondence with the evolutionary status deduced
from a much more detailed spectral analysis (SB00; GD01). The ratio
$x_{INT} / x_{FC}$ can thus be used as an empirical measure of the
time elapsed since the onset of star-formation, allowing us to broadly
assess evolutionary trends in the data.

Fig.~\ref{fig:Evolution} shows that younger systems have stronger
H$\beta$, HeII and [OIII] emission lines, which we tentatively
interpret as evidence for evolutionary effects.  Only composites are
plotted, since only for them we can safely interpret the FC component
as due to young stars and hence read $x_{FC} / x_{INT}$ as an age
indicator. Also, note that $W$'s in this figure are measured with
respect to the whole ``starburst flux'' $F_{SB} = F_{obs} (x_{FC} +
x_{INT})$.  This choice of normalization facilitates the comparison
with spectral evolution models such as Starburst99 (Leitherer \etal
1999), particularly for systems undergoing star-formation over a
period of order $10^8$ yr.  Since H$\beta$ contains a strong
contribution from gas ionized by the starburst,
Fig.~\ref{fig:Evolution}a roughly traces the evolution of the
starburst. The HeII and [OIII] to $F_{SB}$ ratios, on the other hand,
are measures of the relative power of the AGN and the starburst, and
so Fig.~\ref{fig:Evolution}b and c can be used to assess the evolution
of the AGN with respect to that of the starburst.

As two extremes, we consider both an instantaneous burst and constant
star-formation to describe the starburst. In the former case, the
optical continuum due to the young stars peaks in luminosity about 3
Myr after the burst, and then fades by a factor of 3 by 10 Myr and by
another factor of 10 by 50 Myr. The systematic decrease in $W_{HeII}$
and $W_{[OIII]}$ in the more evolved starbursts would therefore require
that the AGN fades more rapidly than this.  In this instantaneous
burst scenario, the AGN lifetime must be very short in the composite
Seyfert 2's ($<$ few $\times 10^7$ years). Since the AGN should
disappear before the intermediate-age stars, the descendants of the
Seyfert 2 nuclei would be part of the post-starburst population (the
``E+A'' galaxies). Thus, the only plausible evolutionary link between
the composite and ``pure'' Seyfert 2's would seemingly require that
latter are the evolutionary {\it precursors} to the former. This does
not seem physically appealing, among other reasons, because the AGN's
in the ``pure'' Seyfert 2's are significantly less powerful on-average
than in their supposed descendants (the youngest composites).
Moreover, an instantaneous burst description is clearly not
appropriate for sources like NGC 5135 and NGC 7130, which contain a
mix of young and intermediate age stars.

Alternatively, we can consider a more long-lived ($\gg 10^7$ years) event
in which the star-formation remains roughly constant.  The $W_{H\beta}$
values predicted by Starburst 99 for constant star formation vary from
150--450 to 10--60 \AA\ between 0 and $10^8$ yr for different initial mass
functions and solar metallicity. This range covers the values in
Fig.~\ref{fig:Evolution}a. The $\sim 10$-fold decrease in $W_{H\beta}$
during this period is also in agreement with the data.  In this model, the
luminosity of the optical continuum steadily grows with time (e.g. by a
factor of 3 to 4 between 10 and 300 Myr after the start of the burst). The
systematic decrease in $W_{HeII}$ and $W_{OIII}$ in the more evolved
starbursts would therefore imply that the AGN's luminosity remains
relatively constant over this time-scale. Long time-scale bursts are indeed
seen in pure starburst galaxies, where ground based spectra usually
integrate over regions containing several young, but non-coeval,
super-associations (Meurer \etal 1995; Gonz\'alez Delgado \etal 1999; Lan\c
con \etal 2001). Individual starburst knots have been identified in some of
our composites (Gonz\'alez Delgado \etal 1998), so we expect this averaging
effect to happen in Seyfert 2's too.  In this scenario, both the starburst
and AGN are relatively long-lived, and the AGN could well outlive the
starburst.  As the star-formation ceases, the starburst will fade first in
H$\beta$, then in the optical continuum, eventually reaching the point
where only scattered FC and AGN-powered lines survive. With the drop in the
starburst-continuum the FC becomes weaker and $W_K$ less diluted. Galaxies
like ESO 362-G8, NGC 5643 and Mrk 78, with their weak FC but pronounced
post-starburst populations may represent this intermediate stage.  [OIII]
and HeII will be much less affected than H$\beta$ by the death of massive
stars, so excitation ratios will go up. Starburst-heated dust emission will
subside, decreasing $L_{FIR}$.  All these suggest that composites will end
up as ``pure'' Seyfert 2's.

Clearly, the above is quite speculative.  Eventually, it may be
possible to test these ideas empirically with improved age dating
techniques which will allow us to evolve detected starbursts back and
forth in time and thus to evaluate whether the age of circum-nuclear
starbursts may be used as a clock for nuclear activity. As in any
other astrophysical context where evolution is important,
statistically significant samples will be necessary to trace all
relevant stages.

\section{Conclusions}

\label{sec:Conclusions}

We have examined a sample of 35 galaxies previously studied in a
series of papers which addressed the occurrence of circum-nuclear
starbursts in Seyfert 2's and implications for the ``starburst-AGN
connection''. The detailed spectral analysis performed in these
previous papers was used to classify the nuclei into 15 composite and
20 ``pure'' Seyfert 2's, with composites defined as nuclei in which
large populations of stars with ages of $10^6$--$10^8$ yr have been
conclusively detected either through UV stellar winds features, high
order Balmer absorption lines in the near UV or the WR bump. Using
this classification as a guide, we investigated continuum colors,
absorption line equivalent widths, emission line properties, far-IR
luminosities and near-UV surface brightnesses for the sample galaxies
and comparison samples of normal and starburst galaxies, with the
primary goal of verifying whether these properties reflect the
presence or otherwise of circum-nuclear starbursts. Since this
consistency was verified, these properties, all of which are more
easily measured than those originally used to detect these starbursts,
offer alternative empirical diagnostics of compositeness which can be
applied to larger and more distant samples.

Our main results can be summarized as follows.

\begin{itemize}

\item[(1)] We have applied a semi-empirical population synthesis
method which effectively provides a bi-parametric description of each
galaxy in terms of old ($\ge 10^9$ yr), post-starburst ($10^8$ yr)
populations and a ``Featureless Continuum'' which represent the
combined contribution of $\le 10^7$ yr stars and a non-stellar FC.  In
practice, this representation requires little more than the equivalent
width of the Ca II K line, and empirical calibrations were presented
which allow an estimation of the ($x_{OLD},x_{INT},x_{FC}$) vector
from $W_K$ and the $F_{3660}/F_{4020}$ color, both easy to measure
observables.  This very simple scheme proved instrumental in unveiling
a number of properties and gives results which are consistent with a
more detailed spectral analysis of the stellar population mixture.

\item[(2)] {\it All} starburst/Seyfert 2 composites but {\it only one}
``pure'' Seyfert 2 (NGC 1068) have $x_{OLD} < 75\%$ at 4861 \AA.  This
threshold can be expressed in a more empirical way by $W_K < 10$
\AA.  Given that our operational definition of composite/``pure''
systems is based on the detection or otherwise of stellar features
from massive stars, the most natural interpretation for this result is
that it is due to a {\it contrast} effect: Seyfert 2's whose
circum-nuclear starbursts are overwhelmed by the light from the old
stars in the host galaxy are not recognized as composite systems. At
least some ``pure'' Seyfert 2's must fit this description, i.e.,
composites with weak starbursts.  This may in fact apply to most, if
not all, ``pure'' Seyfert 2's, since we cannot determine the nature of
their weak UV-excess with the data discussed here.

\item[(3)] A related result is that strong FC sources, those with
$x_{FC} \ga 30\%$, are invariably associated with certified
starburst/Seyfert 2 composites. We have argued that this is expected,
since if scattered light was dominant in such sources then reflected
broad lines would render them classified as type 1 Seyferts.

\item[(4)] The equivalent widths of emission lines (H$\beta$, [OIII],
[OII]), when measured with respect to the FC, are larger in ``pure''
Seyfert 2's than in Starburst galaxies. Composites have $W^{FC}$'s
intermediate between these two extremes.  This fits with the idea that
circum-nuclear starbursts in Seyfert 2's contribute proportionately
more to the FC than to the line flux, since, while the AGN-powered
line emission is seen directly, only a small fraction of the
corresponding non-stellar FC reaches the observer after scattering.

\item[(5)] Part of the separation in $W^{FC}_{H\beta}$ between
``pure'' Seyfert 2's and composites may be due to another contrast
effect, in which the high order Balmer lines, the main diagnostic of
massive stars in the optical, are diluted and eventually filled by
nebular emission in strong lined sources, preventing them from being
recognized as composites.

\item[(6)] Composite systems tend to be less excited in terms of the
HeII/H$\beta$ and [OIII]/H$\beta$ ratios.  This indicates that the
circum-nuclear starbursts have a direct impact on the emission line
properties of Seyfert 2's, with typically $\sim 50\%$ of H$\beta$
being powered by OB stars.

\item[(7)] The dual nature of the ionizing source in composites also
leaves a kinematical imprint on the emission line profiles, which can
be described in terms of a broad component with NLR-like line ratios
and a narrower and (usually) less excited one. These components are
presumably associated with the gas ionized by the AGN and starburst
respectively.

\item[(8)] Composite galaxies are luminous in the far-IR, with
$L_{FIR} \sim 10^{10}$ to $10^{12}$ L$_\odot$ and are more luminous
than ``pure'' Seyfert 2's and normal galaxies. As a whole, composites
are rather similar to merger systems in terms of their far-IR
properties.  About 50\% of the composites, but only 20\% of ``pure''
Seyfert 2's are in interacting systems, suggesting that galaxy
interactions spur circum-nuclear starbursts.

\item[(9)] Composite galaxies have near-UV surface-brightnesses in
their central-most few hundred pc that are on-average an
order-of-magnitude higher than in normal galaxies of any Hubble
type. These high surface brightnesses are similar to those in pure
starburst galaxies.  They are also several times higher on-average
than in the ``pure'' Seyfert 2's (whose central surface brightnesses
in most cases overlap those of normal early-type galaxies).

\item[(10)] These properties were condensed into a set of empirical
criteria which separate composite from ``pure'' systems.  Most of the
diagnostic power resides in the $W_K < 10$ \AA\ limit. Other
properties (small emission line $W^{FC}$, low excitation and large
$L_{FIR}$) serve to confirm the $W_K$-based diagnostic of
compositeness and to judge border line cases. We predict that many
more starburst/Seyfert 2 composites will be uncovered by applying
these diagnostics to large samples of near UV spectra.

\item[(11)] The tendency of composites with older starbursts to have
smaller H$\beta$, HeII and [OIII] emission-line equivalent widths was
tentatively interpreted as a sign of the evolution of circum-nuclear
starbursts in Seyfert 2's. Circum-nuclear starbursts which are
powerful today may fade away and, barring evolution of the active
nucleus, appear as ``pure'' Seyfert 2's.  By improving the dating
techniques and enlarging the databases we hope to trace this
evolutionary sequence and thereby put strong constraints on physical
scenarios for the starburst-AGN connection.

\end{itemize}

\acknowledgments

It is a pleasure to thank the organizers and participants of the
Guilllermo Haro 2000 (INAOE) workshop on ``The Starburst-AGN
connection'', for a stimulating meeting in which this work was
conceived.  We would like to thank R. A. Jansen and C. T. Liu for
making their data available.  RCF thanks the hospitality of Johns
Hopkins University, where this work was developed, and the support
provided by the National Science Foundation through grant \#
GF-1001-99 from the Association of Universities for Research in
Astronomy, Inc., under NSF cooperative agreement AST-9613615. Partial
support from CNPq and PRONEX are also acknowledged.  HRS work was
supported by NASA under Grant No.\ NAG5-9343.  The National Radio
Astronomy Observatory is a facility of the National Science Foundation
operated under cooperative agreement by Associated Universities, Inc.



\end{document}